\documentclass{ar-astro2e}

\usepackage{amsfonts}
\usepackage{amssymb}
\usepackage{amsmath}
\usepackage{graphicx}
\usepackage{ARAstroBib}
\begin{document}

\input epsf.tex    
\input epsf.def   

\input psfig.sty

\jname{..}
\jyear{2000}
\jvol{}
\setcounter{secnumdepth}{4}

\title{Theoretical Challenges in Galaxy Formation}

\markboth{Theoretical Challenges in Galaxy Formation}{Theoretical
  Challenges in Galaxy Formation}

\author{Thorsten Naab$^{1}$ \&  Jeremiah P. Ostriker$^{2,3}$
\affiliation{$^{1}$ Max-Planck-Institute for Astrophysics,
  Karl-Schwarzschild-Str. 1, 85748 Garching, Germany; \\ 
  email: naab@mpa-garching.mpg.de\\$^2$ Department of Astronomy,
  Columbia University, 550 W, 120th Street, New York, NY10027, USA;\\email:jpo@astro.columbia.edu\\ $^3$ Department of Astrophysical
Sciences, Princeton University, Princeton, NJ 08544, USA}}


\begin{abstract}
Numerical simulations have become a major tool for understanding
galaxy formation and evolution. Over the decades the field has
made significant progress. It is now possible to simulate the
formation of individual galaxies and galaxy populations from well
defined initial conditions with realistic abundances and global
properties. An essential component of the calculation is to correctly
estimate the inflow to and outflow from forming galaxies since
observations indicating low formation efficiency and strong
circum-glactic presence of gas are persuasive. Energetic 'feedback'
from massive stars and accreting super-massive 
black holes - generally unresolved in cosmological simulations - plays
a major role for driving galactic outflows, which have been shown to
regulate many aspects of galaxy evolution. A surprisingly large
variety of plausible sub-resolution models succeeds in this 
exercise. They capture the essential characteristics of the
problem, i.e. outflows regulating galactic gas flows, but
their predictive power is limited. In this review we focus on one
major challenge for galaxy formation theory: to understand the
underlying physical processes that regulate the structure of the
interstellar medium, star formation and the driving of galactic
outflows. This requires accurate physical models and numerical
simulations, which can precisely describe the 
multi-phase structure of the interstellar medium on the currently
unresolved few hundred parsecs scales of large scale cosmological
simulations. Such models ultimately require the full accounting for
the dominant cooling and heating processes, the radiation and winds from
massive stars and accreting black holes, an accurate treatment of
supernova explosions as well as the non-thermal components of the
interstellar medium like magnetic fields and cosmic rays.           
\end{abstract}

\maketitle

\section{Introduction}

\subsection{What do we want to learn?}
\label{intro}
Do we understand galaxy formation? Galaxies have been called “the
building blocks of the Universe” and they are clearly the fundamental
units within which stars are organized. They 
do show characteristic sizes ($R_{\mathrm{gal}} \sim kpc$) and
masses ($M_{\mathrm{gal}} \sim 10^{10} M_{\odot}$). Their abundance
($\sim 10^{-2} Mpc^{-3}$) is set by their characteristic mass and
the fact that they constitute a moderate fraction ($f_{\mathrm{gal}}
\sim 10 \%$) of the cosmic baryon budget. Can we derive these numbers
($R_{\mathrm{gal}}, M_{\mathrm{gal}}, f_{\mathrm{gal}}$) from first
principles? Can we, from straightforward numerical simulations, chart
the history of when, where and how the formation and evolution of
galaxies occurred?  And, finally, do we understand it all well enough
to characterize the internal properties of these systems, their ages,
kinematics and mass distributions and their organization into families
having properties describable using relatively few parameters?  

As a problem in physics, there are four clearly definable aspects: (1)
specification of the initial conditions; (2) knowledge of the physical
processes primarily responsible for understanding each phase of 
galactic evolution; (3) computational tools that permit us to start
with (1), utilize (2) to construct models predicting the detailed
properties of representative samples of galaxies to be (4) tested by
direct comparison with the rich treasury of information provided by
nature revealed by modern observational technology. 

We will argue that the current, standard cosmological models are
sufficiently accurate to provide initial conditions as required to any
specified accuracy. With regard to physical processes, the problem is
divided into two parts: (A) what would the evolution be if we only had
to consider dark matter; and (B) how is the picture altered if we
include the primordial radiation fields, the baryonic gas as well as
the energy, processed interstellar matter and momentum input from the stars and
massive black holes? The current state of the art shows a good grasp
of problem (A) - different investigators using different codes recover
quite similar descriptions of the universe; but with regard to the
more complex problem (B) - allowing for 'feedback' from stars and
black holes, we have only preliminary gropings toward physical understanding. One
simple example suffices. The non-thermal and relativistic components
of the interstellar medium - magnetic fields and cosmic rays - are not
thought to be primordial but in our Galaxy have energy densities
comparable with kinetic energy densities, significantly higher than
thermal energy densities 
(e.g. \citealp{1990ApJ...365..544B,2001RvMP...73.1031F}). Are these
components essential to understanding galaxy formation or are they
mere byproducts? They are not included in most treatments, and the
omission may (or may not) be crucial. However, the situation is
improving. 

As a result of our success with regard to problem (A) - the 'stage
setting' so to speak - we have a moderately good grasp of the physics
that determines the approximate values of the three scales, the
numbers for size, mass and abundance, noted as the fundamental
characteristics of galaxies, but we have a poor knowledge of the
details that are important in determining the internal structure
and evolution of these systems.  Finally, our computational tools are
marginally adequate for the simpler part (A) of the task, but perhaps
not up to the challenge of the multi-dimensional, multi-component,
time-dependent computation involving the necessary range of temporal
and spatial scales. 

So far we have been describing this as an {\it ab initio} problem of
physics, like the motion of a playground swing  - though more complex -
but of course this is not the way that the history
unfolded. Observational discoveries have guided us every step of the
way, often pointing out to us how much too simple our models have
been.  These observations have been of two kinds, those that
describe the Universe as it is (e.g. galaxy rotation curves or the
details of the multi-phase interstellar medium) and those
that tell us of the time development, either through using the
archaeological method of examining the stellar populations of nearby
galaxies and determining when/how the various components were
assembled, or by using the Universe as a time machine and looking at
the progenitor populations at earlier cosmic epochs. In any case,
direct observations are the facts that all models have to be tested against
and in most cases they are the drivers for progress
in theoretical galaxy formation, both on the small scales of
individual galaxies as well as on the large scale distribution and
redshift evolution of galaxy populations.  

Galaxy formation has become such a large field in astrophysics that a
full overview of all theoretical challenges is beyond the scope of a
single review. Here we focus on a subset of problems in
computational astrophysics. Numerical implementations of 'feedback' 
processes have traditionally been tested with idealised galaxy models
and merger simulations. These models have resulted in important
insights on star formation, morphological and kinematic 
transformations, merger driven gas flows, triggering of star
formation, the impact of accreting black holes on the termination of
star formation, and size evolution. We give an overview of these
feedback models and their use in modern cosmological simulations of galaxy
formation. We briefly highlight major steps forward like the
successful simulation of spiral galaxies and the cosmological
evolution of galaxy populations. Some other major theoretical challenges
that can be addressed with cosmological galaxy formation simulations are not 
discussed here in detail. One of these is how galaxies accrete
their gas. Gas accretion is a necessary and fundamental process for
galaxy formation but surprisingly it has not yet been conclusively
observed. From numerical studies it is still unclear whether the gas
is accreted onto the galaxies cold filaments \citep{2005MNRAS.363....2K,2009Natur.457..451D} or whether the
filaments dissolve in the halos and accretion is more smooth
\citep{2013MNRAS.429.3353N}. A related question is how galactic outflows actually
transport metal enriched material into the circumgalactic medium
(see e.g. \citealp{2006MNRAS.373.1265O,2015MNRAS.446..521S}). This is
also a numerically challenging question, as the spatial resolution in
the halos of galaxies is typically much lower than in the dense
regions and mixing processes are highly complex
(e.g. \citealp{2015ApJ...805..158S}). We also do not address 
issues that test galaxy formation on small scales in the context of the
underlying cold dark matter cosmological model like the 'Too big to
fail' problem \citep{2011MNRAS.415L..40B} or the question whether small dark matter
halos have cusps or cores
(e.g. \citealp{1998ApJ...502...48K,2012MNRAS.421.3464P}). We accept
teh standard cold dark matter paradigm because of its numerous proven
successes on large scales, while fully aware of the challenges it
faces on small scales. Instead we
focus on the physics of the interstellar medium (ISM). The ISM
strongly influences galaxy formation. Many processes determining star
formation and galactic outflows as well major observable features act
in the ISM and a better understanding and more accurate modelling of
theses processes are, in our view, the major theoretical challenge for
galaxy formation in the future. 

\subsection{Some relevant observations}

Our summary of observations that put important constraints on
theoretical galaxy formation models will necessarily be brief and does
not attempt to provide a full list of references. The first and most
obvious is the observational information accumulated in the last century
specifying the four principle components of all massive galaxies; stars,
gas, dark matter and super-massive black holes: 

\subsubsection{Stars} From Hubble's time onwards we realized that the bulk of the
mass in the visible parts of galaxies resides in one of two
components, a spheroidal part having a scale length typically of only
a few kpc to a few tens of kpc with a roughly de Vaucouleurs’ surface
density profile \citep{1948AnAp...11..247D} and a flattened, rotationally supported disk/spiral
component, which is typically somewhat larger (apart from the highest
mass systems) and has a roughly exponential profile
(e.g. \citealp{2003MNRAS.343..978S,2009ARA&A..47..159B,2011ARA&A..49..301V}). These
two components appear to be distinct, and environmental considerations
must be important in understanding their formation, since isolated
systems tend to be disk dominated and those in regions of high
galactic density tend to be dominated by the spheroidal component
(e.g. \citealp{1980ApJ...236..351D,2009ARA&A..47..159B,2010ApJ...723...54K,2011MNRAS.416.1680C}). Recent
integral field studies have significantly improved our understanding
of the complex kinematics of galaxies
\citep{2002MNRAS.329..513D,2011MNRAS.413..813C,2012A&A...538A...8S,2014MNRAS.443..485F,2015ApJ...798....7B}.
Stellar dating indicates extended and relatively flat star formation
histories for the disks with typical ages of a few billion years and
peaked star formation histories with typical stellar ages of $\sim 10$
billion years for massive early type galaxies
(e.g. \citealp{2003MNRAS.341...33K,2004Natur.428..625H,2005ApJ...621..673T,2006ARA&A..44..141R,2009ApJS..182..216K}). Large
surveys made it possible to observe relatively accurate stellar mass
functions not only in the local Universe
(e.g. \citealp{2008ApJ...675..234P,2009MNRAS.398.2177L,2013MNRAS.436..697B}) 
but also towards higher redshifts
(e.g. \citealp{2012ApJ...754...83B,2013ApJ...767...50M,2013ApJ...777...18M,2014MNRAS.444.2960D,2016ApJ...825....5S}).   

\subsubsection{Gas} Typical Milky Way like spiral galaxies roughly have  $\sim
10 \%$ of their mass in cold interstellar medium gas ($\lesssim 10^4
K$). An even larger fraction (some of it hotter and ionised) gas might be
  stored in the so called 'circum galacitc medium', the region
  extending from the star-forming interstellar medium into the
  galaxies halos (e.g. \citealp{2014ApJ...792....8W,2015ARA&A..53...51S}). More massive early-type
galaxies typically have significantly lower cold gas fractions, although
they are not devoid of cold gas, contrary to the traditional picture
\citep{2011MNRAS.415...32S,2010MNRAS.403..683C,2011MNRAS.414..940Y,2012MNRAS.422.1835S}. Massive
early-type systems are usually embedded in hot ($> 10^5 K$) X-ray  
emitting gas comprising a significant fraction of the total baryonic mass
(e.g. \citealp{2000ARA&A..38..289M,2006ApJ...640..691V,2009ApJ...693.1142S,2009ApJ...703..982G,2010ApJ...719..119D,2012ARA&A..50..353K,2014MNRAS.444.3581R,2015MNRAS.449.3806A}).     

\subsubsection{Dark matter}  Following Zwicky's and Babcock's work in the 1930s and
then the work of many authors on the rotation curves of normal
galaxies in the 1970s and 1980s (e.g. \citealp{2001ARA&A..39..137S}),
it became apparent that the stars in 
most normal galaxies are embedded in massive halos comprised of some
unknown type of dark matter with a total mass and size roughly 10
times that of the stellar component. The generally flat observed
rotation curves of spiral galaxies are an important test for
cosmological formation models (see
\citealp{2014RvMP...86...47C}). Recent result from strong lensing 
have contributed to our knowledge of the dark matter content of
massive galaxies, which have typical contributions of 5\% to 20\%
within their stellar half-light radii
(e.g. \citealp{2009ApJ...703L..51K,2010ARA&A..48...87T}, see also
lensing measurements of the stellar-to-halo mass ratio, \citealp{2006MNRAS.368..715M}). Dwarf
galaxies like Sculptor or Fornax or the recently discovered category
of large utradiffuse galaxies (c.f. \citealp{2015ApJ...798L..45V}) are
dominated by dark matter throughout (see also \citealp{2016ApJ...817...84K}).  

\subsubsection{Super-massive black holes} A number of studies have indicated that super
massive black holes typically reside in the centers of normal galaxies
(having stellar masses $\gtrsim  10^{10.3} M_{\odot}$), with their
masses tightly correlated with the masses (and stellar velocity
dispersions) of the spheroidal components of the galaxies, the ratio
being roughly 5:1000 (see e.g. \citealp{1997MNRAS.291..219G,2013ARA&A..51..511K} for a
review). Given the evident association with AGN, it is widely believed
that the energy emitted by these monsters during their formation is
roughly 10 \% of their rest mass \citep{1973A&A....24..337S,1982MNRAS.200..115S},
that makes them competitive with high mass stars with regard to energy
input (in various forms) into the surrounding galaxies (e.g. \citealp{1998A&A...331L...1S}). \\

\subsubsection{The Milky Way} The archaeological method was used very
successfully in the last half 
of the 20th century to reconstruct a plausible history of our own
galaxy, the Milky Way. The Sun is a typical star in the disk component
that gradually formed from relatively metal rich gas.  It appears that
this disk component grew slowly, in
size and mass, as rotationally supported gas was steadily turned into
stars over cosmic time, and the typical stars in our cosmic
neighborhood were formed only 3-6 billion years ago, relatively late
in the evolution of the Universe. The fact that much less than 10 \%
of the disk stellar mass has a metallicity that is less than 10 
\% of the latest formed stars tells one immediately that the
disk is temporally a 'secondary' structure heavily contaminated by the
metal rich ejecta from earlier stellar generations
\citep{1975ApJ...202..353O}. The age distribution tells us that it
formed 'inside-out' with the stars in the low metallicity, gas rich
outer parts of the disk formed most recently (see
e.g. \citealp{2013A&ARv..21...61R}).  The somewhat tri-axial,
bar-like, inner structure is old and may have formed via the
instability of a cold rotating disk (e.g. \citealp{1973ApJ...186..467O}), 
but the outer spheroidal halo, is likely the debris from in-falling, captured,
smaller systems that has accumulated over time. The stars in this
extended spheroidal (or elliptical) component are typically $\sim 10$
billion years old, are lower in heavy element abundances and tend to
have an isotropic or even somewhat radially biased distribution of
orbits.   

Most of the stars (the fraction might be as high as $\gtrsim 95 \%$)
in our Galaxy were made from gas that was added to the Galaxy, forming
into stars within the system and only a very small fraction of the
stellar mass comes from stars made in other galaxies that were added
to our system via galactic mergers \citep{2012ARA&A..50..531K}. Thus
'major mergers' might not have been at all important in the late
formation history of our Galaxy, or of others with very similar structures. 

Work by \citet{1962ApJ...136..748E} in the early 1960s provided solid
evidence that our galaxy began in ga phase of dramatic 
collapse. Other, spheroidal, systems observed in detail, while more
massive and more metal rich, seemed to be composed of stars of similar
age and orbital properties, so it was plausible that they formed by a
similar process. In this simple picture the disk is a later addition
as higher angular momentum, already contaminated material drifted into
the galaxy, accumulated in a rotating disk and was gradually turned
into the bulk of the stars. This provides a natural explanation for
the two components of the Hubble classification and also a reason for
the absence of the disk components in dense environments within which
tidal or ram-pressure effects prevent the late formation of
disks. While the details of this story have 
evolved, the overall picture has withstood the test of time remarkably
well. The archaeological approach to galaxy formation and evolution
continues, with much useful work being done in teasing out the details
of how the extended spheroidal component was put into place. If this
picture is correct, then in the much more massive elliptical galaxies
like M87 the secondary, stellar component added by the cannibalization
of numerous smaller systems, may comprise 20 \% up to 50 \% of the
total, in contrast to the much smaller fraction of 
accreted stars in the common, lower mass, disk-like spiral systems.

The Milky Way also holds most information about the detailed structure
of the multi-phase interstellar medium (ISM). Most of the gas is found
in three phases, the cold neutral medium, the warm neutral medium and
the hot ionized medium. The hot phase fills about 30\% of the volume
\citep{2001RvMP...73.1031F} in the disk but dominates further than a
few kpc from the disk midplane (see \citealp{2009ARA&A..47...27K}, and
Sec. \ref{ISM}).

\subsection{Learning from galaxy evolution with redshift}
\label{redshift}

Observations of galaxies extending towards higher redshift (and thus
earlier times) have given additional insight in galaxy properties of
fundamental importance.  

\subsubsection{Ubiquitous winds} Galactic winds, with velocities up to
500 $km s^{-1}$ and most likely of bi-conical nature, carrying large
amounts of material out of star forming galaxies (the
rate being comparable to and higher than the star formation rate) are
ubiquitous, not only in the nearby Universe
(e.g. \citealp{1990ApJS...74..833H,1999ApJ...513..156M,2000ApJS..129..493H,2005ARA&A..43..769V,2014ApJ...794..156R})  
but also at higher redshift at the cosmic peak of conversion of gas
into stars \citep{2001ApJ...554..981P, 2010ApJ...717..289S, 2011ARA&A..49..525S,2013ApJ...770...41M}. At low as well
as high redshift these winds most likely enrich the circum-galactic
medium with gas, metals and possibly magnetic fields
\citep{2014ApJ...792....8W,2010ApJ...717..289S,2013ApJ...772L..28B},
providing the material which, if falling back in at later times with
added angular momentum \citep{1969ApJ...155..393P}, can be the source
of the secondary disk systems. These winds transport gas out of
the galaxies at rates similar to which gas is converted into stars and
therefore have to be of importance for regulating the formation
efficiency of stars in galaxies. Even at high redshift the launching
sites of star formation driven \citep{2012ApJ...761...43N} and AGN
driven \citep{2014ApJ...796....7G}  winds can now be resolved with
modern instruments.   

\subsubsection{Size evolution of early-type galaxies} Today's massive ($ \sim
10^{11} M_{\odot}$) early-type galaxies can form early and become 'red
and dead' by $z \sim 2$  as much smaller systems than those seen today
($\sim 1kpc$), with the growth in size (while not forming stars) to be
understood as a likely sign of subsequent addition of stars in minor
mergers at larger radii
(e.g. \citealp{2005ApJ...626..680D,2007MNRAS.382..109T,2010ApJ...709.1018V,2011ApJ...739L..44D}).
Observations of significant structural evolution of massive early-type 
galaxies disfavor any singular monolithic collapse or binary merger
formation scenario \citep{2008ApJ...677L...5V}. Also the observed
strong increase in size and the weak decrease in velocity dispersion
\citep{2009ApJ...696L..43C}  
of the early-type galaxy population as a whole, which also includes
additions to the red sequence at lower redshifts (see
e.g. \citealp{2013ApJ...766...15P,2014ApJ...788...28V,2016arXiv160703493F}),
poses tight constraints on any formation model. From the observed
 age distribution of stars in normal massive early-type galaxies we
 know that the substantial observed evolution was not caused
 primarily by the addition of newly formed stars but rather the
 addition and rearrangement of old stars in these systems.   

\subsubsection{Evolution of spiral galaxies} The high-redshift
progenitors of Milky Way like disk systems are 
also smaller than local examples of similar systems and have formed
half of their mass below $z \sim 1$. Most of the mass is assembling at larger
radii by in-situ star formation providing direct evidence for
'inside-out' growth accompanied by mass growth in the central regions
which can be dominated by bars and bulges \citep{2013ApJ...778..115P}. The
central mass growth might originate from secular instabilities or
merger events, but most stars currently in spiral systems were
  made from gas added to them rather than from accreted stars or
  stellar systems. In general the size evolution of spiral systems is,
however, significantly less rapid than for early-type galaxies
\citep{2014ApJ...788...28V}. 

\subsubsection{Evolution of star formation rates and gas fractions}
\label{eosfr}
A
significant fraction (if not most) of stars in the Universe are formed in 
galaxies with star formation rates that are almost linearly related to
their stellar mass (the ‘star formation main
sequence’) since $z \sim 2.5$  (see
e.g. \citealp{2007ApJ...660L..43N,2007ApJ...670..156D,2012ApJ...754L..29W,2015ApJ...801L..29R}). The tightness
of the overall relation and the mostly disk-like morphology
\citep{2006Natur.442..786G,2009ApJ...706.1364F} of the highly star
forming systems indicates that major merger driven starbursts are of
minor importance for the universal star formation budget. The increase
in star formation rate (the normalization of the ‘main-sequence’)
towards high redshift is accompanied by increasing gas fractions reaching up to
$\sim$ 50 \% at redshift $z \sim 2$
(e.g. \citealp{2010Natur.463..781T,2010ApJ...713..686D,2013ApJ...768...74T}). This
buttresses the simple picture that most star-formation comes from the
gradual transformation of accumulated gas into stars.

\subsection{Methods of solution}

Let us return now to the physics problem to be solved given this
observational background. First we look at what we have described as
part (A) the evolution of radiation fields, dark matter and gas in the
standard cosmological paradigm. This has been well summarized in
several recent textbooks (e.g. \citealp{2010gfe..book.....M}), so only
some of the highlights need to be mentioned. A spectrum of adiabatic
perturbations is imprinted onto the three components at high redshift
producing cosmic microwave background radiation (CBR) fluctuations emitted
at roughly redshift 1000, the analysis of which (cf. WMAP, Planck)
uniquely specifies the cosmological model
\citep{2007ApJS..170..377S,2014A&A...571A..16P}. If we take that to be
the simplest one compatible with the data (the '$\Lambda CDM$' cosmologically flat
model), the model can be defined by five to six independent parameters
that are typically known now (primarily, but not entirely from
analysis of the CBR) to high accuracy. The composition of the dark
matter remains unknown but the standard 'cold dark matter' model has
been so successful that the principle remaining alternatives, Warm
Dark Matter or Fuzzy Dark Matter behave essentially like $\Lambda CDM$
on all large scales with (interesting) deviations becoming apparent
below $\sim 1 kpc$. 

\subsubsection{Direct simulations of dark matter} Accepting this model
we can specify in a cosmologically representative 
volume the statistical distribution of gas, dark matter and radiation
in a fashion sufficiently detailed to provide initial conditions for
computation of the evolution of the various components. In the
simplest treatments of this evolution, where dark matter is
followed via Newton's laws and the transformation of gas into
stars and black holes is ignored. Many different groups have worked on
the problem producing extraordinarily successful (and convergent)
results (see \citealp{2012AnP...524..507F}). The Millennium simulation
\citep{2005Natur.435..629S} was perhaps the most publicly successful
such dark matter calculation, but other simulations
\citep{2011ApJ...740..102K} also  of  larger volumes
\citep{2012MNRAS.426.2046A} or constrained to a certain halo mass
scale \citep{2007ApJ...667..859D,
  2008MNRAS.391.1685S,2009MNRAS.398L..21S,2012MNRAS.425.2169G}  have
made very important contributions. This is problem (A) and it is
essentially a solved problem. But it leaves us a long way from
understanding the evolution of real galaxies composed primarily of
stars.  

\subsubsection{Semi-analytical models for baryons} There exist different approaches
to the more difficult part (B), the allowance for star and black hole
formation and the input from these sources 
of mass, energy, momentum and processed matter back into the gaseous
component. The first approach to this hard problem was to set up
comprehensible 'model problems' the solutions of which would be
illuminating. One large class of such efforts has been broadly labeled
the 'semi-analytic' method, where one takes the dark matter
simulations as a given, and then tries, by one means or another, to
estimate how the other components will react. Examples of progress
made in the late 90's via the setting and
solving of very informative 'model problems' consider the formation of
disks from gas accumulating within dark matter halos
\citep{1997ApJ...482..659D,1998MNRAS.295..319M}. Modern attempts to
input what are thought to be the most important physical processes in
a simple fashion
(e.g. \citealp{1993MNRAS.264..201K,1999MNRAS.310.1087S,2006MNRAS.365...11C,2006MNRAS.370..645B,2011MNRAS.413..101G,2015MNRAS.451.2663H})
aim at finding that set which best produces realistic mock 
observations (see review by \citealp{2015ARA&A..53...51S}). Another
class of more analytical models makes the simplifying assumption that
star forming galaxies evolve in a quasi-equilibrium fashion regulated
by gas inflow and outflow, star formation and the change of mass in
the galactic gas reservoir. The above approaches are extensively
reviewed in \citet{2015ARA&A..53...51S}. 

\subsubsection{Direct simulations including baryons} While these
methods have been most helpful in furthering our
understanding, the technical and algorithmic progress has enabled the
direct and ambitious effort to include as much of the detailed physics
as possible and simply compute forwards from the well established
initial conditions to the current time using gravity, hydrodynamics,
radiation transfer and all of the elaborate apparatus developed by
physics to address continuum mechanics.  The computational tools to
follow the evolution of dark matter and stars (gravity) as well as gas
(hydrodynamics) have been developed since the early 1980’s.   

The first three-dimensional coupled hydrodynamical simulations
including self-gravity used the smoothed particle hydrodynamics (SPH)
technique
\citep{1981MNRAS.194..503E,1988MNRAS.235..911E,1989ApJS...70..419H}. The
Lagrangian particle based SPH method
(\citet{1977MNRAS.181..375G,1977AJ.....82.1013L}, see also
\citet{2010ARA&A..48..391S} and \citet{2015ARA&A..53...51S} for recent   
reviews) is relatively simple to implement and due to its adaptive
spatial resolution and good conservation properties has been very
popular for galaxy formation simulations until today.  However, the
basic implementation has to be modified for typical astrophysical
conditions including shocks, shear and large temperature
gradients and it has become clear that some ‘standard’
implementations have serious difficulties to properly model fluid
mixing and sub-sonic turbulence
\citep{2007MNRAS.380..963A,2010ARA&A..48..391S}. Most of the recent
SPH work on cosmological galaxy formation is based on derivatives of
either the \textit{GASOLINE} \citep{2004NewA....9..137W} code or the
\textit{GADGET} \citep{2005MNRAS.364.1105S} code and include updated
implementations to treat the mixing problem better (see
e.g. \citealp{2008MNRAS.387..427W,2012MNRAS.422.3037R,2014MNRAS.443.1173H,2014MNRAS.445..581H,2015MNRAS.446..521S,2015MNRAS.454.2277S}
and references therein). As an alternative to the SPH method particle
based meshless-finite-mass and meshless-finite-volume methods have
been proposed \citep{2011MNRAS.414..129G}. The recent \textit{GIZMO}
implementation is based on the \textit{GADGET} framework and shows
some significant improvements on idealised test problems, in
particular for low Mach number gas \citep{2015MNRAS.450...53H}.  

 Eulerian hydrodynamic codes have also been 
widely used for cosmological simulations, some with adaptive mesh
refinement capabilities. These codes typically perform better than SPH
in terms of mixing and shock problems but might suffer from artifacts due to
grid structure and numerical diffusion, which, for some solvers, can
become significant. The first rough Eulerian treatment was by
\citet{1992ApJ...393...22C} and the recently most used, greatly
improved, Eulerian adaptive mesh refinement codes are \textit{ENZO}
\citep{2014ApJS..211...19B}, \textit{RAMSES} 
\citep{2002A&A...385..337T}, and ART \citep{1997ApJS..111...73K} and
also \textit{FLASH} \citep{2000ApJS..131..273F} as well as
\textit{ATHENA} \citep{2008ApJS..178..137S} for ISM simulations on
smaller scales. The newly developed moving mesh code \textit{AREPO}
\citep{2010MNRAS.401..791S} similarly suffers from
numerical diffusion but combines advantages of the Lagrangian and  
Eulerian approaches and performs much better than traditional SPH codes
like \textit{GADGET} on mixing problems with a high convergence rate
\citep{2010ARA&A..48..391S,2012MNRAS.424.2999S}.

There are ongoing
efforts to better understand the strengths and weaknesses of different
numerical schemes 
(e.g. \citealp{2010MNRAS.406.1659P,2011MNRAS.415..271H,2013MNRAS.432..711H,2014MNRAS.442.1992H,2014ApJS..210...14K})
and to constantly improve on accuracy and performance of all major codes. It has been realized
early on that different numerical schemes applied to cosmological
simulations can result in systems with different physical properties
\citep{1999ApJ...525..554F}, even if only gravity and hydrodynamics
are considered.  In addition, there is a wealth of published sub-resolution
models (see Section \ref{stellarfeedback}) which are used to model
galaxy formation. These models are often designed for particular numerical
schemes and introduce even stronger variations in physical properties
for a given set of initial conditions \citep{2012MNRAS.423.1726S}.
One of the major challenges in computational galaxy formation is to
further improve on the numerical schemes and reduce the contribution
of sub-resolution modeling to numerically resolved physical scenarios.

These numerical methods, which we could label 'ab initio' computations
aiming to solve part (B), will be discussed in the second part of this
review. But first we will address two other extremely useful idealized and
empirical approaches which preceded and accompanied them.  

\subsection{Disks, Ellipticals and Mergers - a very useful set of
  idealized simulations} 
\label{mergers}
In the early 70s a definite cosmological model had not emerged and the
computational resources as well as the numerical algorithms were still
limited. This was the start of idealized merger simulations as it
had been realized that galaxies actually interact and merge for 
bridges, tidal tails and other merger phenomena are observed
\citep{1972ApJ...178..623T,1985MNRAS.214...87J,1988ApJ...325...74S}. Early self-consistent N-body simulations
(e.g. \citealp{1978MNRAS.184..185W}) were limited to the stellar
component of galaxies with a few hundred gravitating particles
(stars), a situation that has significantly improved until now when
millions of star particles, dark  matter particles and complicated gas
dynamical processes can be studied
(e.g. \citealp{2013MNRAS.433...78H}). With methods for
creating equilibrium models for multi-component galaxies
(i.e. \citealp{1993ApJS...86..389H}) it became possible to simulate
the evolution of the stellar and gaseous components of disk galaxies
in more detail. For more than 20 years such setups are playing a
major role for developing star formation and feedback models in direct
comparison with observations of star forming spiral galaxies 
(e.g. \citealp{1994ApJ...437..611M,2003MNRAS.339..289S,2005ApJ...620L..19L,2008MNRAS.387.1431D,2011MNRAS.417..950H,2013ApJ...770...25A}).
In most cases it is tested under which conditions a given model
reproduces the observed relation between gas surface 
density and star formation rate surface density \citep{1998ApJ...498..541K}. Only
successful models are then considered for more complex simulations of
galaxy mergers or the cosmological formation of galaxies.

A number of important physical processes have been investigated with
merger simulations and  
the insight into galaxy formation physics has been significant. We know
that equal-mass mergers are rare and relatively unimportant for the
cosmic star formation budget 
(see Sec. \ref{eosfr} and \citealp{2012ApJ...744...85M,2009MNRAS.394L..51B,2011ApJ...738L..25W}. For 
intermediate mass galaxies (e.g. Milky Way) and low mass systems stars
are primarily formed from streams of gas that accumulate centrally or
in disks (e.g. \citealp{2016MNRAS.458.2371R,2017MNRAS.464.1659Q}). For
high mass systems (i.e. massive early-type galaxies) mergers become
important. Stars added in major and minor mergers can make up as much as 50 \% of
the largely outer envelopes (in case of minor mergers) of these systems
\citep{2009ApJ...699L.178N,2016MNRAS.458.2371R,2017MNRAS.464.1659Q}. But the initial
proposal that normal ellipticals are made by morphological
transformations of disk galaxies in binary major mergers of
spirals, though not generally applicable \citep{1980ComAp...8..177O},
is influential and instructive. In particular, it was shown that
observed nearby disk galaxy mergers most likely evolve into systems
with structural similarities to young early-type galaxies (e.g. \citealp{2004AJ....128.2098R}).  

\subsubsection{Collisionless mergers} Merger simulations might be
separated into two groups. Collisionless 
simulations of stars and dark matter mutually interacting
by gravity alone were evolved by the collsionless Boltzmann
equation (see e.g. \citealp{2008gady.book.....B}). Such idealized systems can be 
considered energy conserving (no radiative losses). In reality almost
no galactic system meets these conditions. Even massive galaxies have
some amount of hot and cold gas
(e.g. \citealp{2009ApJ...693.1142S,2011MNRAS.414..940Y,2012MNRAS.422.1835S}). But
if the gas components can be considered dynamically unimportant it is
justified to consider a system collisionless (the term 'dry' 
has been used in the literature). When spheroidal one-component systems
merge, their structural evolution can -to good accuracy - be estimated
using the virial theorem with only a few 
assumptions \citep{2000MNRAS.319..168C,2009ApJ...699L.178N,2009ApJ...697.1290B}.
Following \citet{2009ApJ...699L.178N} one can assume that a compact
initial stellar system has formed (e.g. involving gas dissipation)
with  a total energy $E_i$, a mass $M_i$, a gravitational radius
$r_{g,i}$, and the mean  square speed of the stars is $\langle
v_i^2\rangle$. According to the virial theorem
\citep{2008gady.book.....B} the total energy of the system is   

\begin{eqnarray} 
E_i & = &  K_i+W_i = -K_i = \frac{1}{2} W_i \nonumber \\
    & = & -\frac{1}{2} M_i \langle v_i^2 \rangle = -\frac{1}{2} \frac{GM_i^2}{r_{g,i}}.
\label{virial}
\end{eqnarray}

This system then merges (on zero energy orbits) with other systems of
a total energy $E_a$, total mass  $M_a$, gravitational radii
$r_{a}$ and mean square speeds averaging $\langle  v_a^2\rangle$. The
fractional mass increase from all the merged galaxies is $\eta =
M_a/M_i$ and the total kinetic energy of the material is $K_a=(1/2)
M_a \langle v_a^2\rangle$, further defining  $\epsilon = \langle v_a^2
\rangle/\langle v_i^2\rangle $. Here $\epsilon = 1$ represents an equal mass merger and $\epsilon \sim 0$ for very minor mergers. Under the assumption of energy
conservation (e.g. \citet{2006A&A...445..403K}
indicate that most dark matter halos merge on parabolic orbits) the
ratio of initial to final mean square speeds, gravitational radii and
densities can be expressed as 

\begin{equation}
\frac{\langle v_f^2\rangle }{\langle v_i^2\rangle } =
\frac{(1+\eta\epsilon)}{1+\eta} \nonumber \\ , 
\frac{r_{g,f}}{r_{g,i}} = \frac{(1+\eta)^2}{(1+\eta\epsilon)}
\nonumber \\ ,
\frac{\rho_f}{\rho_i} = \frac{(1+\eta \epsilon)^3}{(1+\eta)^5}.
\end{equation}

For binary mergers of identical systems, $\eta = 1$, the mean square
speed remains unchanged, the size increases by a factor of two
and the densities decrease by a factor of four. In the limit that the
mass is accreted in the form of a weakly bound stellar systems with 
$\langle v_a^2\rangle  << \langle v_i^2\rangle $ or $\epsilon << 1$,
the mean square speed is reduced by a factor two, the size increases
by a factor four and the density drops by a factor of 32. These
estimates are, however, idealized assuming one-component systems, no
violent relaxation and zero-energy orbits with fixed angular
momentum. In the presence of a dark matter halo the structural changes
become more complicated and e.g. the fraction of dark matter at the
center (inside the half-mass radius of the stars) may increase due to
violent relaxation \citep{2005MNRAS.362..184B,2012MNRAS.425.3119H}. 
In general a fraction of the orbital angular momentum of the galaxies
will be transferred to rotation in the central galaxy, so that the merger
remnants in most cases rotate significantly
\citep{1979ApJ...229L...9W,2006ApJ...636L..81N,2009A&A...501L...9D,2010MNRAS.406.2405B,2011MNRAS.416.1654B}.   
Major mergers of spheroidal galaxies are also expected to flatten
existing abundance gradients \citep{1979MNRAS.189..831W,2009A&A...499..427D}.

If a spheroidal system experiences collisionless minor mergers (with
satellite galaxies of much lower mass than the central) violent
relaxation effects in the central galaxy are negligible and the
satellite stars are stripped at larger
radii \citep{1983MNRAS.204..219V}, a mechanism that offers a
plausible explanation for the observed structural evolution of massive
galaxies \citep{2003MNRAS.342..501N,2012MNRAS.425.3119H}  and the
formation of extended stellar envelopes in early-type galaxies leading
to the very high observed Sersic indices  and outer metallicity
gradients \citep{1983MNRAS.204..219V,2013MNRAS.429.2924H}. Whether
minor mergers alone can explain the observed strong size evolution
of massive early-type galaxies will depend on the actual merger rates as
well as the structure of the satellite galaxies
\citep{2009ApJ...706L..86N,2012MNRAS.422L..62C,2012ApJ...746..162N,2013MNRAS.428..641O,2013MNRAS.431..767B}. 
Per added unit of stellar mass this process can also increase the
fraction of dark matter within a half-mass radius more efficiently
than major mergers \citep{2005MNRAS.362..184B,2013MNRAS.429.2924H}. 

Another important process investigated is the morphological transformation
of kinematically cold disk galaxies to kinematically hot spheroidal
galaxies \citep{1982ApJ...259..103F,1983MNRAS.205.1009N,1992ARA&A..30..705B,1992ApJ...393..484B,1992ApJ...400..460H}. 
Violent relaxation heats the disk stars and some fraction of the
orbital angular momentum and of the spin of the initial disk systems can be
absorbed by the dark matter halos \citep{1988ApJ...331..699B}. This
results in stellar remnants that can have early-type galaxy morphology
and kinematics if the progenitor galaxies had a bulge component of
sufficiently high phase space density
\citep{1993ApJ...409..548H}. Therefore merging is important for the
formatin of hot stellar systems. Depending on the mass-ratio of the
merging disks - and the amount of 'damage' that is done to the primary
disk, the remnants rotate fast or slow, have disky, round or boxy
isophotal shapes and are more or less flattened
\citep{1998giis.conf..275B,1998ApJ...502L.133B,1994ApJ...427..165H,1999ApJ...523L.133N,2000MNRAS.316..315B,2001ApJ...554..291C,2003ApJ...597..893N,2004A&A...418L..27B,2005A&A...437...69B,2005MNRAS.357..753G,2009MNRAS.397.1202J}. 
For very low mass infalling systems the disk might be only moderately
heated and retains its flat and rotationally supported morphology for
single events
\citep{1993ApJ...403...74Q,1999MNRAS.304..254V}. Repeated minor
mergers will make the initial disk system more spherical and reduce 
its spin \citep{2007A&A...476.1179B,2010A&A...515A..11Q}.   

\subsubsection{Mergers with gas}
A major step in understanding galaxy mergers was established once the
simulations included a dissipative gas component. The
gravitational torques exerted on the gas during the merger were able 
to drive the gas from large radii to the nuclear regions of the merger
remnant once it lost its rotational support in tidally induced
shocks \citep{1996ApJ...471..115B}. This has important implications
for galaxy formation. Using  sub-resolution models for the
conversion of gas into stars (see Section \ref{starformation}), 
it was shown by many studies that the gas inflow can trigger a nuclear
starburst similar to what is observed in ultra-luminous infrared
galaxies (ULIRGS) and explain observations of 'extra light' in the
centers of low mass early-type galaxies
\citep{1994ApJ...437L..47M,1994ApJ...431L...9M,1996ApJ...464..641M,1999ASPC..182..124K,2000MNRAS.312..859S,2009ApJS..181..135H,2009ApJS..181..486H,2009ApJS..182..216K,2010ApJ...720L.149T,2013MNRAS.433...78H,2014MNRAS.442.1992H}.
Gas accumulating at the center of merger remnants also makes the potential
more spherical, favoring the population of stars on tube orbits
\citep{1996ApJ...471..115B,2007MNRAS.376..997J}. As a result, rotating
remnants of gas rich mergers can form disk-like subsystems
\citep{1997ApJ...478L..17B,2000MNRAS.316..315B,2002MNRAS.333..481B,2007MNRAS.376..997J,2009MNRAS.397.1202J}
and show observed line-of-sight velocity distributions with steep
leading wings, which is not the case if gas is neglected
\citep{2006MNRAS.372..839N,2009ApJ...705..920H,2010ApJ...723..818H}.   

The effect of dissipation in binary galaxy mergers has also been used
to explain the detailed shape of scaling relations and the fundamental plane and its potential evolution with redshift
\citep{2006MNRAS.373.1013C,2006MNRAS.370.1445D,2006ApJ...650..791C,2006ApJ...641...21R,2009ApJ...691.1424H}. One
branch of binary merger simulations focused on the potential feeding 
of supermassive black holes, which are observed in most nearby
early-type galaxies. Here the merger-triggered inflow provides the low
angular momentum gas to be accreted onto the black hole
\citep{1989Natur.340..687H}. The energy released from the accreting
black hole, on the other hand, has been suggested 
to drive gas out of the merger remnant, significantly reducing its
star formation rate \citep{2005ApJ...620L..79S}. The idea of  'black
hole feedback' to 'quench massive galaxies' was born. Assuming a
Bondi-like accretion and a relatively simple scaling for energy
feedback in a sub-resolution model , it
was also possible to provide an explanation for the observed 
stellar mass black hole mass relation
\citep{2005Natur.433..604D}. This finding has led to a number of
studies, based on idealized binary merger simulations investigating
the evolution of the $M_{\mathrm{bulge}}-M_{\mathrm{BH}}$ and
  $\sigma_{\mathrm{bulge}}-M_{\mathrm{BH}}$ 
relation
(e.g. \citealp{2006ApJ...641...90R,2009ApJ...690..802J,2010MNRAS.406L..55D,2011MNRAS.412.1341D,2014MNRAS.442..440C,2014MNRAS.437.1456B}) 
and the evolution of the quasar luminosity function based on the
assumption that most of the AGN activity is driven by galaxy mergers
(e.g. \citealp{2005ApJ...630..705H,2006ApJS..163....1H}).    

Despite the explanatory successes of these studies, the drawback of
idealized sub-resolution models used is that the actual physical processes
(e.g. the feedback from central super-massive black holes) cannot be
resolved and the cosmological context is omitted. This is true for
most merger simulations even though the typical spatial and mass resolution
is much higher than in larger scale cosmological simulations (Section
\ref{abinitio}). The assumptions are mostly simple and physical effects are
condensed or hidden in parameters or scale factors, which are often
scaled to a specific set of observations (see Section
\ref{intro}). Therefore the validity of the astrophysical
implications always remains somewhat uncertain. For example many
models adopted in binary merger simulations use a simple accretion scheme to determine
the mass flow rates onto the black holes. The accretion rate depends
on the sound-speed of the surrounding medium, which can vary significantly depending on the
assumed star formation and feedback model (see Section
\ref{blackholefeedback}). Alternative models link the 
accretion rates to gravitational instabilities and torques  
\citep{2011MNRAS.415.1027H,2012MNRAS.420..320H}. Some models used for
binary merger simulations assume feedback in the form of thermal energy to
the gas surrounding the black hole \citep{2005ApJ...620L..79S}, others
take into account the observed momentum output which significantly
reduces the amount of hot coronal gas and the observable X-ray
luminosities \citep{2011MNRAS.412.1341D,2014MNRAS.442..440C}. Binary
merger experiments have been used as test beds for 
sub-resolution models used in larger scale cosmological simulations.

\begin{figure}
  \centering
      \includegraphics[width=1\linewidth]{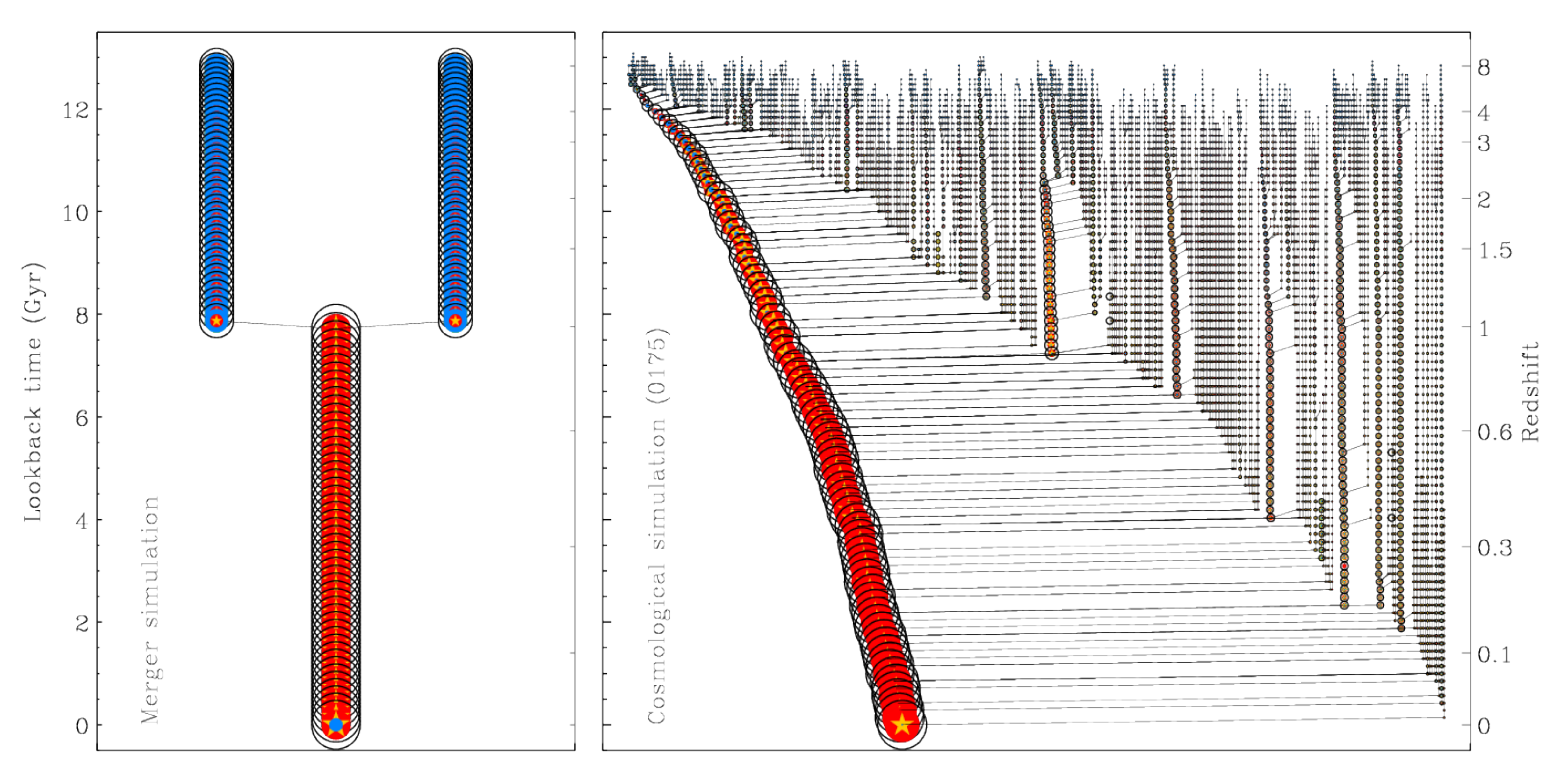}
\caption{Binary disk merger simulations are useful in understanding
  merging disk galaxies observed in the sky. In general, they lack the realism and
  complexity of the cosmological assembly of old massive early-type galaxies. In
  the left panel we show a schematic and very simplified binary disk 'merger-tree'. Two gas-rich (blue) stellar
    (yellow) disks with little hot gas (red) merge at $z \approx 1$ and
    form an elliptical galaxy. In a cosmological zoom simulation
    (right panel) of the formation of a dark matter halo (black circles) and its massive galaxy
    (cold gas: blue) is significantly more complex. It is evident that
    continuous infall of matter in small and large units is an
    important characteristic of the assembly of massive galaxies. The
    galaxy shown (0175 from \citealp{2010ApJ...725.2312O}) is another
    extreme case as it has no major merger since $z \approx 3$. Others
    galaxies of similar mass can have up to three major mergers. Major
    mergers definitely  happen and they have a strong impact on galaxy
    evolution. Cosmological assembly and mass growth, however, is
    always accompanied by numerous minor mergers and gas accretion
    (figure from \citealp{2014MNRAS.444.3357N}).}
\label{tree}
\end{figure}

\subsubsection{Caveats of the merger hypothesis}
The importance of major mergers for the formation and evolution of 
massive galaxies is still under debate. Idealized merger simulations
ignore the cosmological context where gas accretion, repeated minor
mergers as well as environmental effects are important. As the
expected major merger rates are low some massive galaxies might experience no
major merger at all (see Fig. \ref{tree}). It is clear that galaxy
mergers, in particular of equal mass, can have a significant impact on galaxy
kinematics and mass growth if they happen at late times. The
importance of stars formed in merger-triggered starbursts may have
been overestimated as well, in particular due to the fact that most merger
simulations ignored halo gas accretion. If halo accretion is included
the disks have more realistic constant star formation rates and the
contribution from merger triggered star formation is significantly
lower and sometimes negligible \citep{2011MNRAS.415.3750M}. This is
supported by observations indicating that most star formation in the
Universe happens in relatively ‘normal’ morphologically settled disk-like
galaxies (e.g. \citealp{2007ApJ...670..156D}). Merging systems with
enhanced star formation rates seem to be of minor importance but might
help in the transition to quenched early-type galaxies
\citep{2011ApJ...742...96W}. Observations at low and high redshift also  
provide evidence that significant black hole accretion is not solely
connected to merging but also gas rich disk like galaxies can host AGN
of significant luminosity
\citep{2009MNRAS.397..623G,2011ApJ...726...57C,2011ApJ...727L..31S,2012ApJ...744..148K}. Also, cooling flow instabilities within the hot gas of elliptical galaxies lead to a ‘secular branch’ of AGN fueling.  Apparently
major mergers can host luminous AGN but by no means are all AGN induced by 
mergers.  Also it seems unlikely that the  
population of present day early type galaxies can have directly
formed via mergers among the population of present day disk galaxies
and their progenitors, as the early-type population is too old, too massive
and too metal rich \citep{2009ApJ...690.1452N}. At earlier times the
discrepancy in mass and size between observed Milky Way progenitors and
massive early-type galaxies is even more pronounced \citep{2013ApJ...778..115P}.

\subsection{Ranking and Matching}
\label{matching}

A fundamental question in galaxy formation, embedded in the modern
hierarchical cold dark matter framework, is how much of the available
baryonic matter is converted into stars in the central galaxies in
dark matter halos. This quantity might be termed galaxy formation
efficiency or galaxy fraction $f_{\mathrm{gal}}$. There have been a
number of attempts to estimate this number for the Milky Way. Whereas
the stellar mass of the Milky Way is relatively well determined
\citep{2013A&ARv..21...61R}, the major uncertainty is in the mass of
the Milky Way's dark matter halo. Typically mass estimates are in the
range of $1- 2 \times 10^{12} M_{\odot}$
\citep{2008ApJ...684.1143X,2008MNRAS.384.1459L,2012ApJ...759..131B,2010MNRAS.406..264W,2016ARA&A..54..529B}. 
The masses result in galaxy fractions of $f_{\mathrm{gal}} \sim 20 -
40 \%$. With much better observed stellar mass functions at low and
high redshifts and converged dark matter simulations for a 
given cosmological model, it has become possible to estimate the galaxy
formation efficiency (or the relation of galaxy mass to halo mass) for
a large range of halo masses locally and at higher redshifts. The methods used
include halo occupation distribution modeling
\citep{2002ApJ...575..587B,2002MNRAS.329..246B,2004ApJ...609...35K}, 
conditional luminosity function modeling \citep{2003MNRAS.339.1057Y}
or a rank ordered matching of observed galaxy mass functions to
simulated halo mass functions
\citep{2004MNRAS.353..189V,2006MNRAS.371.1173V,2006ApJ...643...14S,2006ApJ...647..201C,2010ApJ...717..379B,2010ApJ...710..903M,2013MNRAS.428.3121M,2013ApJ...770...57B,2013MNRAS.435.1313H,2015MNRAS.450.1604L}.   

Most of these studies indicate that around 10 - 20 \% of the
available baryons are converted into stars in dark matter halos of
$\sim 10^{12} M_{\odot}$. This fraction is lower in dark matter halos
of higher and lower mass with considerable uncertainties at both ends
(see
e.g. \citealp{2013ApJ...778...14G,2014arXiv1401.7329K,2014MNRAS.437.3228G}). The
mismatch of most early cosmological simulations with these 
empirical estimates was highlighted in 
\citet{2010MNRAS.404.1111G} and galaxy fractions became a standard test
presented in almost every publication about cosmological
simulations. Also at higher redshift the tension with simulations was formerly
much more severe (see e.g. \citealp{2013MNRAS.428.3121M}) due to
the overly efficient early conversion of baryons into stars.  

The matching models also provide an independent estimate
of the amount of stars formed in the galaxies (in-situ star formation
as measured by the star formation rates) and the amount of stars
accreted in galaxy mergers. The general conclusion is that all
galaxies are dominated by in-situ star formation at high redshift ($
z\gtrsim 1.5$). A trend that continues to low redshift for moderate mass
(Milky Way type) galaxies which are predicted to have accreted between
5 \% \citep{2013MNRAS.428.3121M} and 30 \%
\citep{2013ApJ...770...57B} of their stellar mass. High mass galaxies
assemble more and more of their stellar mass by mergers towards lower
redshifts. However, the estimated fractions of accreted stars by z=0
of galaxies in massive halos ($M_{\mathrm{halo}} \sim 10^{13}$) vary
significantly between 20\% and 60 \%
 \citep{2013MNRAS.428.3121M,2013ApJ...770...57B,2013ApJ...770..115Y}.    
The general trend is similar to simulations
(e.g. \citealp{2010ApJ...725.2312O,2012MNRAS.425..641L,2012MNRAS.427.1816G,2016MNRAS.458.2371R,2017MNRAS.464.1659Q}).   
It has also been highlighted that galaxies in massive halos ($M_{halo}
\gtrsim 10^{13} M_{\odot}$) form their stars before the
halo assembles. Low mass galaxies form their stellar components after
their halos assemble \citep{2009ApJ...696..620C}.

\section{Ab initio simulations of galaxy formation}
\label{abinitio}

\subsection{Star formation and gas cooling} 
\label{starformation}

Most modern cosmological galaxy formation simulations allow for metal
enrichment and metal dependent radiation equilibrium cooling (for specific implementations
see e.g. \citealp{2003ApJ...590L...1K,2005MNRAS.364..552S,2006MNRAS.373.1265O,2007MNRAS.382.1050T,2009MNRAS.393...99W,2009ApJ...695..292C,2013MNRAS.436.3031V})  
of gas in the presence of the UV/X-ray background radiation from quasars and galaxies
(e.g. \citealp{2009ApJ...703.1416F,2012ApJ...746..125H}). If the gas is
cooling rapidly, or in the presence of a rapidly changing radiation field,
non-equilibrium cooling will be more accurate (see
e.g. \citealp{2007ApJS..168..213G,2013MNRAS.434.1063O}) and first
steps in this direction have been made in galaxy scale simulations  
\citep{2013MNRAS.434.1043O,2014MNRAS.442.2780R,2016MNRAS.458.3528H,2016MNRAS.458..270R,2016Natur.535..523F,2016MNRAS.460.2157O}. 
Out of the cool gas reservoir the formation of the stellar populations
is modeled in a simplified way as the relevant spatial and temporal
scales as well as the complex ISM physics cannot be resolved in a
cosmological context. 

Cosmological simulations typically treat star formation in a Schmidt-type manner
\citep{1959ApJ...129..243S} relating the local star formation rate
density to the gas density divided by a time-scale. This time-scale
depends on local gas properties like the dynamical and/or gas cooling
time as introduced by \citet{1992ApJ...391..502K} and
\citet{1992ApJ...399L.113C}. Some implementations couple the star
formation rate to molecular gas  properties inspired by observed
connections
\citep{2002ApJ...569..157W,2007ApJ...671..333K,2008AJ....136.2782L,2008AJ....136.2846B,2012ARA&A..50..531K}  
and use a constant time-scale in combination with the variable local $H_2$
fraction
(e.g. \citealp{2006ApJ...645.1024P,2008ApJ...680.1083R,2009ApJ...697...55G,2012MNRAS.421.2485M,2012MNRAS.425.3058C,2012ApJ...747..124F,2012ApJ...749...36K}) 
which is, however, itself connected to the $H_2$ formation
time-scale. It should be noted that $H_2$ based star formation models
in galaxy simulations add another level of complexity and
uncertainty. The small-scale structure of the ISM, the detailed radiation
field, ionization degrees, magnetic field strengths, all relevant for 
$H_2$ formation
(e.g. \citealp{2007ApJ...659.1317G,2014A&A...570A..81H,2015MNRAS.454..238W}), are
unresolved in most galaxy scale and all cosmological simulations (see,
however, \citealp{2012MNRAS.421.3488H}) and it is unclear whether $H_2$
formation is the primary driver for star formation (see
e.g. \citealp{2012MNRAS.421....9G,2013MNRAS.436.2747K}).          

The star formation models typically require a normalization, the star
formation efficiency, as well as a parameter determining the scaling
with gas density. The parameters are adjusted to match the
zero point and the slope of the observed 
Kennicutt-Schmidt relation between star formation rate surface density
and gas surface density \citep{1998ApJ...498..541K}. The basic
implementations have been extended by sub-resolution models to capture
some characteristics of the multi-phase structure of the gas
(e.g. \citealp{2003MNRAS.339..289S}).  An alternative implementation
for star formation is based on gas pressure (see
e.g. \citealp{2008MNRAS.383.1210S,2015MNRAS.446..521S}) assuming that
galatic disks are in approximate vertical pressure equilibrium. Such a
model has been shown to be in good agreement with the observed
Kennicutt-Schmidt relation with no need for additional calibration
\citep{2008MNRAS.383.1210S}. If used in combination with a fixed
equation of state for the star forming gas the behaviour is similar to
a density dependent criterion as the relation of gas 
pressure and density is fixed. In general, only gas below a 
certain temperature and above a certain density, which can be metal
dependent, is eligible for star formation (see
\citet{2013MNRAS.432.2647H} for a discussion of star formation
criteria). Also in large scale cosmological simulations these
parameters  must be calibrated as the ISM is in general unresolved
and gas cooling is effectively not followed below a few thousand Kelvin
\citep{2015MNRAS.446..521S,2014MNRAS.444.1518V,2015MNRAS.450.1349K}. 

\subsection{The formation of disk dominated systems}

Even though the formation of Milky Way systems has turned out to be
the more difficult problem, it was historically the 
first one tackled. Can we make the Milky Way from reasonable
cosmological initial conditions applying the relevant physical
processes? Simulators typically have used the 'zoom technique'
\citep{1994MNRAS.267..401N} wherein a representative region of the
Universe is simulated first with a dark matter only code, and 
then a relevant high density piece is re-simulated with a
hydro-dynamical code with allowance for the gravitational forces and
gas inflow due to the surrounding matter.    

In hierarchical cosmological models for the formation of galaxies,
small structures form first, grow, and merge into larger objects. In
this framework, galaxies form through the cooling of gas at the
centers of dark matter halos, where it condenses into stars
\citep{1978MNRAS.183..341W}. To match the observed properties of
galaxies and galaxy clusters, purely gravitational processes on their
own cannot account for cosmological structure formation, but gas
cooling/dissipation processes must be considered
\citep{1977ApJ...211..638S,1977MNRAS.179..541R,1977ApJ...215..483B}. These
three very early papers already presented the physical arguments for
the observed scales of galaxies noted in our first paragraph. It had 
been realized early-on from analytical estimates that the conservation
of angular momentum of the cooling gas within dark matter halos could
lead to the formation of galactic disks with flat rotation curves
\citep{1980MNRAS.193..189F}. In early work \citep{1978MNRAS.183..341W}
it was already noted that at high-redshift gas has to be prevented
from excessive cooling into overly dense regions - possibly by
feedback from massive stars to avoid the overproduction of condensed
baryonic matter 
\citep{1974MNRAS.169..229L,1986ApJ...303...39D, 1991ApJ...380..320N}.  
Also, to produce dynamically cold and thin stellar, extended disks, the
accretion of high angular momentum gas from outer regions of the
halos is needed in the more recent past
\citep{1979Natur.281..200F}. This would require feedback
processes to eject gas and avoid early over-efficient star formation
at high redshift as well as  the formation of gas reservoirs to allow
the gas to return at low redshifts with higher angular momentum.  

Early cosmological simulations including the dissipative gas
component (but neglecting star formation) confirmed the problem
\citep{1991ApJ...380..320N,1991ApJ...377..365K,1994MNRAS.267..401N}. Too
many baryons settled into disks which were much more compact than
observed spiral galaxies with too high rotation velocities due to
substantial angular momentum loss during the assembly process caused
by mergers. Not only was the angular momentum for the forming gaseous
disks too low, but also too many baryons would be locked up  
in galaxies \citep{1995MNRAS.275...56N,1997ApJ...478...13N}. 

The over-cooling problem was confirmed by many studies that
followed (see e.g. \citealp{2001MNRAS.326.1228B}). Once a stellar
component was included in the simulation it was possible to
approximately treat the feedback from young stellar
populations. In addition to radiative cooling the role of energy
injection by supernovae could be tested. Investigators quickly
discovered that, while the detailed implementation can change the 
results significantly
(e.g. \citealp{1993MNRAS.265..271N,1999ApJ...519..501S,2000ApJ...545..728T,2004ApJ...606...32R,2005MNRAS.363.1299O})
almost all cosmological simulations resulted in the overproduction of
stars, in low angular momentum bulges
(e.g. \citealp{1996ApJS..105...19K,2001MNRAS.326.1228B,2005MNRAS.363....2K}). It
was again suggested that the origin of the problem lies at   
higher redshift
\citep{2002ApJ...576...21V,2004ApJ...612L..13D}. Galaxies would be 
less concentrated and have higher specific angular momentum if gas
cooling were suppressed before the host halo has assembled
\citep{1998MNRAS.300..773W}.  Still, simulations experimenting with
thermal and kinetic energy injection 
(e.g. \citealp{2003ApJ...591..499A} and many that followed) resulted
in similar problems, with the conclusion that the assumed feedback
models were insufficient to prevent the early collapse of low angular
momentum baryons and their conversion into stars, a problem that
remained unsolved for a long time.

Thus, the early attempts at ab
initio cosmological computations failed or only partially succeeded to
make disk systems that were as low mass and extended in space and time
of formation as those in the real Universe
\citep{2003ApJ...591..499A,2003ApJ...597...21A,2004ApJ...606...32R,2004ApJ...607..688G,2006ApJ...645..986R,2009MNRAS.396..696S,2011MNRAS.410.2625P,2011MNRAS.410.1391A}.
It is important to recall that observations had shown (see Section
\ref{intro}) that real, forming galaxies were embedded in strong
gaseous outflows that were missing from the simulations. 

\begin{figure}
\centering
\includegraphics[width=1\linewidth]{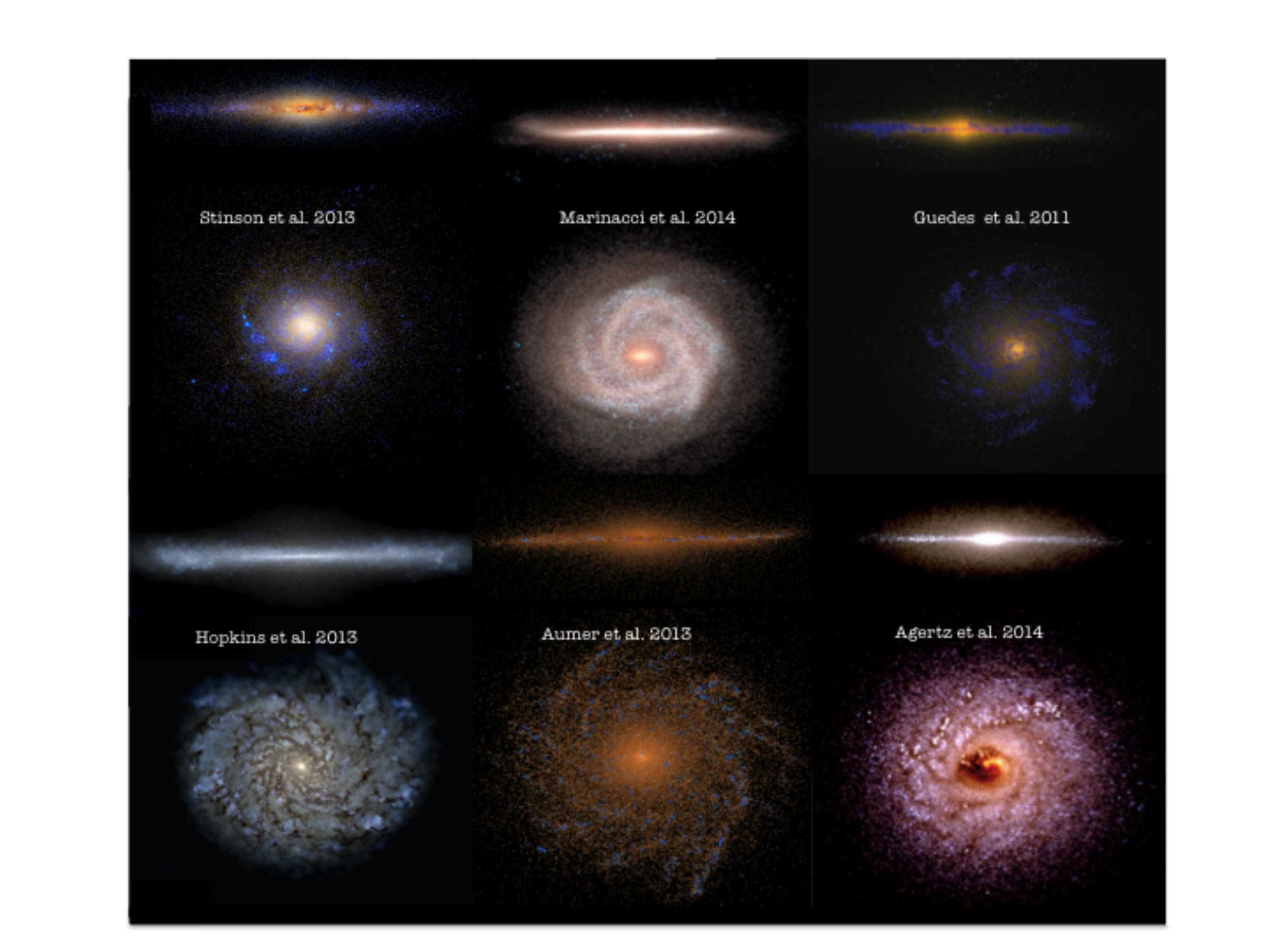}
\caption{Recent cosmological zoom simulations with strong stellar
  feedback of galaxies with sprial like morphologies. The pictures
  show mock images of the stellar light. {\it{Top left}}: SPH (\textit{GASOLINE}) simulation
  of \citet{2013MNRAS.428..129S} including dust attenuation. {\it{Top
      middle}}: Moving-mesh (\textit{AREPO}) simulation of \citet{2014MNRAS.437.1750M}.
{\it{Top right}}: SPH (\textit{GASOLINE}) simulation of
\citet{2011ApJ...742...76G}. {\it{Bottom left}}:  SPH (\textit{GASOLINE})
simulation of \citet{2014MNRAS.445..581H}. Only the face-on view includes dust
attenuation. {\it{Bottom middle}}: SPH (\textit{GADGET}) simulation of
\citet{2013MNRAS.434.3142A}. {\it{Bottom right}}: AMR (\textit{RAMSES}) simulation 
  of \citet{2015ApJ...804...18A}. Only the face-on view includes dust
  attenuation.} 
\label{disks}
\end{figure}

Recently, a number of groups have made significant progress on
reducing the galaxy fraction, $f_{\mathrm{gal}}$, in halos of $\sim
10^{12} M_{\odot}$ and at the same time forming spiral galaxies with
more realistic properties. In Fig. \ref{disks} we  
show examples from six groups who recently succeeded in producing disks with
spiral like morphologies but very different
simulation codes. These studies utilize a variety of qualitatively
different sub-resolution approaches to the response of the 
high-density star forming gas on newly formed and dying stellar
populations. All these 'successful' approaches have in common that gas in 
dense star forming regions can efficiently be pushed out in a galactic
outflow and, possibly, escape from the galaxies and their dark matter
halos. With outflow launching, Milky Way like halos develop 
disk-like galaxies. Detailed investigations of gas flows in the
forming galaxies - which are most easily followed in Lagrangian SPH 
simulations (see also \citealp{2013MNRAS.435.1426G,2015MNRAS.448...59N,2016MNRAS.460.2881N})- have
revealed most characteristics and consequences of galactic outflows. With
strong stellar 
feedback, a significant fraction of the low angular momentum gas
cooling to the centers of dark matter halos at high redshift is
prevented from being converted into stars and can be blown out of the
galaxies. When the proto-galaxies are still small and have shallow
potential wells, this gas will leave the galaxies and never return or
can return at much later times with angular momentum enhanced by
non-linear gravitational torques or mixing
(e.g. \citealp{2011MNRAS.415.1534M}) with the rotating halo gas  
\citep{2011MNRAS.415.1051B,2014MNRAS.443.2092U}. The outflow 
supresses the formation of stellar bulges from the low angular momentum
gas at high redshift and enriches the circum-galactic medium with
metals
\citep{2010Natur.463..203G,2012MNRAS.419..771B,2014MNRAS.442.3745M,2016ApJ...824...57C}. It
also reduces the  
previously reported dramatic effects of mergers. The gas still looses
angular momentum but a significant fraction can be 
ejected before stars are formed. This is particularly efficient if the   
mergers happen early in smaller proto-galaxies \citep{2014MNRAS.443.2092U}.  

Contrary to what has been believed for a long time, even galaxies with
early major mergers can evolve into present day disk-like galaxies with 
low bulge fractions \citep{2005ApJ...622L...9S,2006ApJ...645..986R,2014MNRAS.441.3679A}.
For more massive systems the enriched gas is kept within the halo (in
a galactic fountain) and is accreted back onto the galaxy later on with
metallicity enhanced, sometimes repeatedly
\citep{2008MNRAS.387..577O,2010MNRAS.406.2325O,2012A&A...540A..56P,2014MNRAS.443.3809B,2015ApJ...804L..40G}. 
This process reduces star formation and delays the onset of galaxy 
formation in halos of all masses, in much better agreement with high
redshift abundance matching constraints
(e.g. \citealp{2013MNRAS.428..129S,2013MNRAS.436.2929H,2014MNRAS.445..581H}).
The late accretion of gas with high angular momentum from outside the
halo is increased as fewer baryons were converted into stars in
accreted structures. This is pretty much as it was predicted 38 years ago
\citep{1979Natur.281..200F}. 

\begin{figure}
\centering
  \includegraphics[width=1\linewidth]{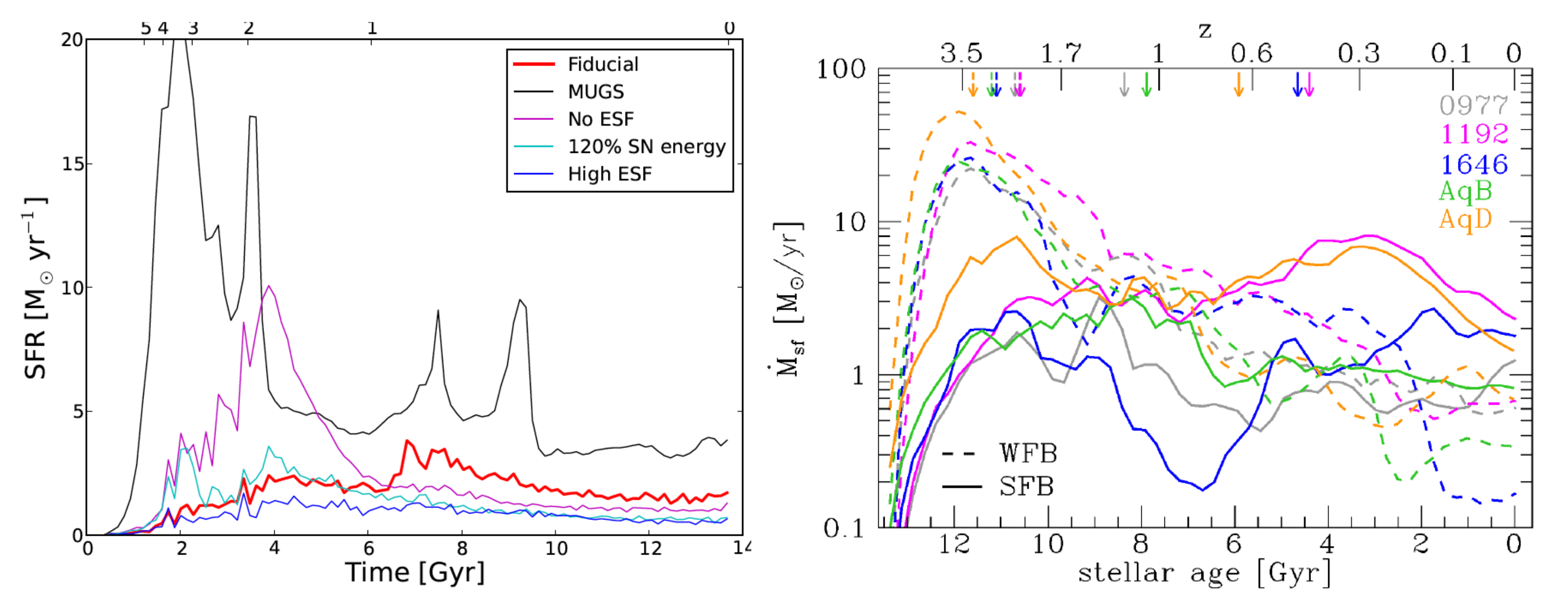}
  
  \caption{The effect of stellar feedback on the star formation
  histories of simulated disk galaxies. Stronger feedback results in
  the suppression of early star formation, relatively flat star
  formation histories and, in the cases shown here, disk like
  morphologies. The flatter star formation rate histories but not the disk-like
  morphologies are generic for all simulations with strong
  feedback. Flat star formation histories are in better agreement with
  observations of Milky Way-sized spiral galaxies. {\it{Left 
      panel}}: The simulations presented   in
  \citet{2013MNRAS.428..129S} show peaked star formation histories 
  for simulations with weak feedback (MUGS
  \citealp{2010MNRAS.408..812S}, No ESF). Star formation at high
  redshift is  suppressed for models with strong feedback (Fiducial,
  120\% SN energy, High ESF). {\it{Right panel}}: The same trend is seen
  in a comparison of five galaxies simulated with weak (dashed lines,
  \citealp{2010ApJ...725.2312O}) and strong feedback (solid lines,
  \citealp{2013MNRAS.434.3142A}). The stellar half-mass formation
  times (arrows) are shifted from $z\sim 2$ to $z \lesssim 1$
  \citep{2014MNRAS.443.2092U}.}    
\label{sfhist}
\end{figure}

In addition, at low redshift a moderately constant gas accretion
rate onto spiral galaxies can be sustained by enriched gas that has been
cycling within the halo of the galaxy
\citep{2010MNRAS.406.2325O,2014MNRAS.443.2092U,2016ApJ...824...57C}. Therefore gas
accretion onto the galaxy is decoupled from the halo assembly,
resulting in flatter star formation histories more consistent with
observations
\citep{2013MNRAS.436.2929H,2013MNRAS.428..129S,2014MNRAS.445..581H,2014MNRAS.442..732W}. 
In Fig. \ref{sfhist} we show two recent examples of the generic effect
of strong feedback on galactic star formation rate histories. Additionally,
Milky Way progenitor galaxies are larger (with strong feedback models)
already at high redshift  and the overall cosmological evolution in
size is significantly reduced
\citep{2013MNRAS.436.2929H,2014MNRAS.441.3679A} in much better 
agreement with observations. The above results rely on 'recipes' to
treat physical processes that are relevant below the resolution scale
of the simulations and sometimes even impact the simulation in well
resolved regions. The variety of these models is remarkable and a good
understanding of the strengths and weaknesses of these models is
relevant to assess whether scientific progress has been made or
whether the good agreement with observations is the result of an
empirical matching exercise. In the following we review the general
pathways followed by different groups.

\subsubsection {Current sub-resolution models for feedback from stellar populations}
\label{stellarfeedback}

Feedback from massive stars has long been suggested to resolve the over-cooling
problem in galaxy formation and various different sub-resolution
feedback models have been presented to approximate the complex
physical processes. Here we call a sub-resolution model an empirical,
physically motivated numerical recipe representing the large scale
impact of energy, mass and momentum during the life and death of massive stars
in state of the art cosmological simulations of large volumes. These
models are necessary as the finest resolution elements (in large scale
cosmological simulations) are typically a few hundred parsecs which
makes it impossible for these simulations to capture the small-scale
multi-phase structure of the galactic ISM. In Section \ref{supernovae}
we will demonstrate why these models can only be a crude
representation of reality, limiting the predictive power of present
day galaxy formation simulations.    

One class of models might be termed 'delayed cooling' models
\citep{1997PhDT........19G,2000ApJ...545..728T}. In one incarnation 
of this approach \citep{2006MNRAS.373.1074S} the energy from
supernova explosions is injected into neighboring gas but the
cooling is  'turned off' for gas inside the expected Sedov blast wave radius
\citep{1977ApJ...218..148M}. This way gas can efficiently be
heated and accelerated. Although often being criticized as being
unphysical due to the suppression of gas cooling, the model attempts to
allow for formation of super-bubbles
(e.g. \citealp{1988ApJ...324..776M}, see \citet{2015MNRAS.453.3499K}
for a modern implementation of superbubble formation in cosmological
simulations and Section \ref{supernovae}). The 'delayed cooling’ model
significantly reduces the galaxy stellar masses and promotes the formation of
disk dominated systems \citep{2007MNRAS.374.1479G,2011ApJ...742...76G,2016MNRAS.tmp.1145K}. 
The models have been extended, in a simplified way, by taking into account the
additional energy release from massive stars before they explode as
supernova
\citep{2013MNRAS.428..129S,2014MNRAS.437.2882K,2015MNRAS.454...83W}. It
has been pointed out by \citet{2016arXiv160901296R} that the delayed
cooling approach results in a significant amount of thermally unstable
circumgalctic gas, which can be problematic when estimating the
gas emission and absorption in galactic halos.

A related approach is 'stochastic thermal' feedback
\citep{2012MNRAS.426..140D}, which does not suffer from some of the
problems of the 'delayed cooling' approach (see
e.g. \citealp{2016arXiv160901296R}). Here the mean thermal energy
injection per unit formed stellar mass is fixed and neighboring 
star particles are, stochastically, only heated if their
temperature can be moved above a certain temperature threshold
(e.g. $\gtrsim 10^{7.5} K$). This guarantees long cooling times, the
onset of a Sedov phase and efficient momentum generation similar to
the 'delayed cooling' models. The total energy injection is an adjustable
parameter and can slighly exceed the available supernova energy
\citep{2015MNRAS.450.1937C,2015MNRAS.446..521S} compensating for the
artificial overcooling. However, the model has been demonstrated to drive strong winds
\citep{2012MNRAS.426..140D}. It has been used for one of the most
successful cosmological galaxy formation simulation suite (the Eagle
simulations, \citealp{2015MNRAS.446..521S}) in terms of
matching observed galaxy population properties and their evolution
with redshift
\citep{2015MNRAS.446..521S,2015MNRAS.450.4486F,2015MNRAS.452.2034R,2016MNRAS.456.1115B}.    

The gas cooling is also delayed in 'non-thermal' heating
models \citep{2013MNRAS.429.3068T}. Here the energy is injected into a
non-thermal energy component, representing turbulence, magnetic fields
or cosmic rays, with a dissipation time-scale of $\sim 10 Myr$. The
energy injection procedure follows the 'stochastic thermal' heating 
(e.g. \citealp{2014MNRAS.444.2837R}). 

A 'two phase' approach is followed in \citet{2006MNRAS.371.1125S} where
the hot and the cold gas phase are evolved separately using SPH
\citep{2003MNRAS.345..561M}. The supernova energy added to a cold gas
particle is stored (i.e. decoupled from the hydrodynamics) and only 
released when it can become a constituent of the hot phase. Accounting
for the momentum input by supernovae and (potential) radiation pressure in a
simplified way this model also produces spiral galaxies with realistic 
properties (see Fig. \ref{disks}, \citealp{2013MNRAS.434.3142A}).   

Another popular approach might be termed 'wind feedback'. Some
fraction of the energy released by massive stars is injected into the
surrounding gas in the form of energy or momentum by which it is
driven away from the region of star formation. The ‘wind’ is
parameterized by a mass loading factor $\eta$, i.e. the ratio of wind
mass-loss to the star formation rate and by a wind velocity
$v_{\mathrm{wind}}$. The original implementation assumed a constant
mass loading and a fixed wind velocity coupled to a stiff effective
equation of state for the gas resulting from thermal energy input from
supernovae \citep{2003MNRAS.339..289S}. Here the gas in the wind is
decoupled from the hydodynamical calculations when leaving the star
forming reagions with its given velocity and is later (the conditions
depend on the respective implementations) incorporated in
the calculations again (see also
\citealp{2014MNRAS.444.1518V}). Observations and  theoretical
considerations, however, indicate decreasing mass-loading and 
increasing wind velocities in higher-mass galaxies with higher star
formation rates \citep{2005ApJ...621..227M,2005ApJ...618..569M}. This
motivated \citet{2006MNRAS.373.1265O,2008MNRAS.387..577O} to introduce
a momentum driven wind model. The wind velocity scales with the velocity
dispersion of stars in the galaxies (or the dark matter,
\citealp{2013MNRAS.436.3031V}) $v_{\mathrm{wind}} \propto \sigma$,
the momentum input scales with the star formation rate, $\dot{m}_{\mathrm{wind}} \times
v_{\mathrm{wind}} \propto \dot{m}_*$, and the mass loading is inversely
proportional to the velocity dispersion $\eta \propto
dm_{\mathrm{wind}}/dm_* \propto 1/v_{\mathrm{wind}} \propto
1/\sigma_*$. Again, to ensure that the gas leaves the star-forming
regions it is then decoupled and later re-incorporated in regions with
lower gas density.  

Although this model has become popular, some authors
\citep{2010MNRAS.402.1536S} turned off the 'wind decoupling' 
the consequences of which are discussed in
\citet{2008MNRAS.387.1431D}. Although empirical in nature, the
galactic wind models results in realistic (compared to observations)
enrichment histories of galaxies and the circum-galactic medium
\citep{2008MNRAS.385.2181F,2011MNRAS.416.1354D}, lower conversion
efficiencies in particular in the regime of disk galaxies
\citep{2010MNRAS.409.1541S,2013MNRAS.428.2966P}, reasonable abundances and flatter star
formation histories for low mass galaxies and higher gas fraction for
star forming galaxies at high redshift
\citep{2008MNRAS.387..577O,2011MNRAS.416.1354D,2013MNRAS.436.2929H} 
and a more realistic cosmic star formation history \citep{2010MNRAS.402.1536S}. Although
originally developed for SPH simulations, decoupled 'momentum driven'
winds have also been used in recent moving mesh simulations with wind
velocities scaled to the local dark matter velocity dispersions (e.g. the
Illustris simulation, \citealp{2014MNRAS.444.1518V}). With cosmological
zoom simulations it has been demonstrated that realistic present
day spiral galaxies \citep{2014MNRAS.437.1750M,2016MNRAS.459..199G} as
well as gas rich massive high-redshift disks
\citep{2012ApJ...745...11G,2014ApJ...782...84A} can be formed with a 
momentum driven wind model in which 'decoupling' has been applied.

Alternatively, an energy driven wind model has been proposed
\citep{2010MNRAS.406..208O} to explain the low abundance of satellite
galaxies in the Milky Way. Here the wind velocity also is assumed to
scale with the velocity dispersion $v_{\mathrm{wind}} \propto \sigma$,
the energy input scales with the star formation rate,
$\dot{m}_{\mathrm{wind}} \times v_{\mathrm{wind}}^{2} \propto \dot{m}_*$, and
the mass loading is taken to be inversely proportional to the square
of the stellar velocity dispersion $\eta \propto dm_{\mathrm{wind}}/dm_*
\propto 1/v_{\mathrm{wind}}^{2} \propto 1/\sigma^2$. This model
results in the same wind speeds but higher mass-loading for lower mass
galaxies and better agreement for the Milky Way satellites luminosity
function \citep{2010MNRAS.406..208O}. In a hybrid model 
\citet{2013MNRAS.434.2645D} combine a momentum driven wind scaling 
with the energy driven wind scaling for galaxies below $\sigma = 75 km
s^{-1}$ to obtain a better match to the galaxy mass function at low
masses (see also \citet{2013MNRAS.430.3213B} for a 'radially varying
wind model'). This transition was motivated by the idea that low mass 
galaxies are more affected by supernova explosions whereas the effect
of radiation pressure takes over at higher masses
\citep{2005ApJ...618..569M,2010ApJ...709..191M}. In an updated
incarnation of the 'decoupled' wind model \citet{2016arXiv160401418D}
have used scaling from high-resolution cosmological zoom simulations
\citep{2015MNRAS.454.2691M} to set the mass-loading and wind
velocities. Efforts are underway to replace these heuristic methods
with others based more closely on high-resolution multi-phase
physical modeling.

\subsection{The formation of bulge dominated systems}  
\label{blackholefeedback}

Spheroidal early-type galaxies have been 'easier' to simulate from
straightforward cosmological initial conditions (any cosmological
simulation with weak feedback will result in the overproduction of
spheroidal galaxies). The escape velocities 
are larger for these more massive systems and so the energy input from
feedback matters somewhat less, and, while still important, stellar feedback
appears to be less critical to the formation process. Empirically they
are known to form in dense regions starting at early times and the
observed structures of proto-ellipticals are quite small as seen at
redshift z = 2-3 (see Section \ref{redshift}). Thus, the difficulties
encountered in making disk-like systems - too early star formation in
too concentrated systems - are alleviated for the construction of
physically plausible spheroidal systems. As a consequence,
cosmological simulations with weak stellar feedback and without AGN
feedback have effectively been used as reasonable 
initial models for the formation of massive, early type galaxies. In
these simulations the final galaxies follow observed early-type galaxy
scaling relations of size and velocity dispersion with stellar mass
\citep{2007ApJ...658..710N,2010ApJ...725.2312O,2010ApJ...709..218F,2012ApJ...754..115J,2011ApJ...736...88F}. They
also have plausible stellar populations with metallicity distributions
that are modulated by their merger history
\citep{2004MNRAS.347..740K,2005MNRAS.361.1216K}. However, 
compared to observations based on abundance matching constraints, 
their stellar masses are about a factor of 2 - 4 too high at a given
halo mass \citep{2010ApJ...725.2312O}. This has been
presumably due to the lack of sufficient energy input (see
e.g. \citealp{2003ApJ...590..619M,2006MNRAS.365...11C}), and it was
alleviated as AGN feedback simulations have been improved.     

For high halo masses, the evolution of the galaxies shows a clear 
two-phase characteristic
\citep{2007ApJ...658..710N,2010ApJ...725.2312O,2010ApJ...709..218F,2012ApJ...754..115J,2013MNRAS.436.3507N,2016MNRAS.458.2371R,2017MNRAS.464.1659Q}.
At early times ($z \gtrsim 1.5$) the galaxies grow by 
in-situ star formation in the deep potential wells of massive
halos. As the low angular momentum gas is efficiently converted into
stars, some of the systems can be remarkably small (also supported by
mergers, \citealp{2010ApJ...722.1666W,2011ApJ...730....4B}), very similar to
the population of observed high-redshift compact galaxies
\citep{2010ApJ...721.1755S,2010ApJ...725.2312O,2012ApJ...744...63O,2016MNRAS.456.1030W}.
Towards lower redshifts in-situ star formation becomes less 
important as cold gas can no longer easily penetrate the 
shocked hot gaseous halos (e.g. \citealp{2003MNRAS.345..349B}) and the
mass assembly becomes dominated by the accretion of stars that have
formed in other galaxies. Cosmological simulations clearly indicate that stellar
accretion is more important at higher galaxy masses
\citep{2010ApJ...725.2312O,2012MNRAS.425..641L,2012MNRAS.427.1816G,2016MNRAS.458.2371R,2017MNRAS.464.1659Q}. This
robust trend is also found in 
abundance matching estimates (Section \ref{matching}) and semi-analytical 
galaxy formation models
\citep{2006MNRAS.370..902K,2008MNRAS.384....2G} and it is strongest for
central galaxies in galaxy clusters
\citep{1977ApJ...217L.125O,2007MNRAS.375....2D}. The late,
collisionless assembly has important consequences for the structural
evolution of the system. As a significant fraction of mass can be
accreted in mergers with smaller and less bound systems
\citep{2012MNRAS.427.1816G,2012MNRAS.425..641L}, this stellar mass is 
added to the systems at large radii
\citep{2010ApJ...725.2312O,2013MNRAS.436.3507N,2016MNRAS.458.2371R,2017MNRAS.464.1659Q}. The resulting strong
increase in galaxy size is driven by  accreted stars.  Simple
arguments based on the virial theorem
\citep{2009ApJ...699L.178N,2009ApJ...697.1290B} show that the same  
mass added in many minor mergers will produce a much more extended
galaxy than if that mass had been added in fewer, more major
mergers (Eq. \ref{virial}, Section \ref{mergers}) . Together with the
weak decrease in velocity dispersion the evolution of the individual
model galaxies is consistent with observational estimates.   

The low present day star formation rates and spheroidal shapes of
galaxies simulated with weak stellar feedback are primarily caused by
the efficient early gas depletion and early conversion of gas into
stars in combination with efficient shock heating of the halo gas and
gravitational heating caused by the accretion of smaller systems
\citep{2009ApJ...697L..38J,2008ApJ...680...54K}. Still, the weak
feedback models provide an attractive start for the physical solution
of the observed structural evolution of massive galaxies (see also
\citealp{2011ApJ...736...88F,2014arXiv1404.3212F}). In its 
extreme limit the assembly of brightest cluster galaxies and the size
evolution of cluster galaxies can be well explained in a substantially
collisionless cosmological assembly model assuming that all stars in
cluster progenitor galaxies have formed before $z \sim 2$
\citep{2013MNRAS.435..901L}.  

With simulations neglecting AGN feedback it has been shown that
the formation history of massive galaxies leaves its imprint on the gas
and stellar kinematic properties of present day early-type galaxies
\citep{2014MNRAS.444.3357N,2014MNRAS.438.2701W,2014MNRAS.444.3388S}. Epochs
dominated by gas dissipation will result in the formation of flattened
stellar distributions (disks), supported by rotation. Major mergers
are rare, and during minor merger dominated phases the stellar systems
experience stripping and violent relaxation, existing cold gas may be
driven to the central regions causing starbursts, trigger the
formation and growth of super-massive black holes or be expelled from
the systems in galactic winds. This, in turn, will impact the
distribution of cold gas and the kinematics of stars forming
thereof. With improved cosmological simulations we move towards a
better understanding of the angular momentum evolution of galaxies
(see \citealp{2015ApJ...804L..40G,2016MNRAS.460.4466Z}). In a first step, 
using cosmological zoom simulations 
\citep{2014MNRAS.444.3357N} have been able to  demonstrate that gas
dissipation and merging result in observable features (at present day)
in the two-dimensional kinematic properties of galaxies, which are
clear signatures of distinct formation processes.

Many of these
features are in agreement with all the valuable predictions 
from isolated merger simulations (see Section
\ref{mergers}). Dissipation favors the formation of fast rotating
systems and line-of-sight velocity distributions with steep leading
wings, a property that can be directly traced back to the orbital
composition of the systems (\citealp{2014MNRAS.445.1065R}, see also
\citet{2012MNRAS.422.1863B} for the obital distribution of dark
matter). Merging and 
accretion can result in fast or  slowly rotating systems with
counter-rotating cores, cold nuclear or extended (sometimes
counter-rotating) disks showing dumbbell-like features – all observed
in real galaxies and in part well understood from binary merger
experiments.     

We have seen that models with stronger feedback and metal cooling,
applied to cosmological simulations, delay the onset of star formation
in more massive halos and the systems become more gas rich
at high redshift, a trend that continues towards low redshift. This
makes galaxies too massive with too high star formation rates in
particular at the central regions 
(e.g. \citealp{2012ARA&A..50..353K}). This had also been found prior to
cosmological simulations (e.g. \citealp{1991ApJ...376..380C,1997ApJ...487L.105C}), due 
simply to the inevitable cooling flows occurring in massive
systems. One dimensional and two dimensional high resolution
simulations of the effect of AGN feedback
\citep{1995MNRAS.276..663B,2001ApJ...551..131C,2005MNRAS.358..168S,2011ApJ...737...26N}
indicated that AGN feedback alleviates  this problem. With
ab inito cosmological simulations a number of groups have now demonstrated
that 'feedback' from an accreting supermassive black hole can suppress
the residual star formation in the central regions of massive
galaxies, confirming the proposal put forward by
\citet{1998A&A...331L...1S}. In the following we review some
sub-resoluton approaches for implementing feedback from supermassive
black holes. 

\subsubsection{Current models for feedback from super-massive black holes}
\label{backholes}

Several different models have been proposed to approximate the effect
of AGN feedback and to follow it in cosmological simulations. In most
galaxy scale sub-resolution models (starting from
\citealp{2005MNRAS.361..776S}) the accretion rate onto the black 
hole $\dot{M}_{\mathrm{BH}}$ is computed based on the
Bondi-Hoyle-Lyttleton formula (actually invented for spherical
accretion of interstellar gas onto the Sun,
\citealp{1939PCPS...35..405H,1944MNRAS.104..273B,1952MNRAS.112..195B})  

\begin{eqnarray}
\frac{dM_{BH}}{dt} = \alpha_{\mathrm{boost}} \frac{4 \pi G^2 M_{BH}^2 \rho}{(c_s^2 +v_{\mathrm{rel}}^2)^{3/2}}.
\end{eqnarray}

Here $c_s$ is the sound speed of the surrounding gas and
$v_{\mathrm{rel}}$ is the relative velocity of the black hole and the gas (see
\citet{2015MNRAS.454.1038R} for modifications of the Bondi rate due to an
assumed viscous accretion disk, see also \citet{2016MNRAS.458..816H}
for the possible failure of Bondi accretion in high-resolution simulations). Also included
in many models is an adjustable accretion 'boost factor'
$\alpha_{\mathrm{boost}}$ which can have values up to 100 in some
implementations. The general motivation for using a boost factor is
the low resolution of the simulations which are unable to follow the
accurate multi-phase gas structure near the black holes and therefore the
accretion rates (see discussion in \citealp{2009MNRAS.398...53B}). In
fact many implementations use Bondi accretion in combination with a
stiff equation of state for the high density gas resulting in gas with
increasing temperature at higher densities - contrary to the expected
structure in the dense ISM   
\citep{2010MNRAS.406..822M,2013MNRAS.428.2966P,2013MNRAS.436.3031V,2014MNRAS.444.1518V,2014MNRAS.442.2304H,2015MNRAS.450.1349K,2015MNRAS.446..521S}. As 
a result the sound speed becomes artificially high and a high 
$\alpha_{\mathrm{boost}}$ compensates for this. From a practical, not
physical, point of view the boost factor ensures that enough gas is
accreted to grow super-massive black holes of reasonable masses. As in
\citet{2005MNRAS.361..776S}, the accretion in most models is limited by
the Eddington rate. In some recent implementations (see,
however, \citealp{2007ApJ...665..107P}) the relative velocity between
the black hole and the ambient medium is not considered (the Lyttleton
part) as the black holes are continuously centered to the potential
minimum of the host halo
(e.g. \citealp{2013MNRAS.428.2966P,2014MNRAS.444.1518V}). \citet{2012ApJ...754..125C} 
take an alternate approach that does not have a boost factor; they
stochastically treat the overlap of the smoothing sphere and the Bondi
sphere, thereby statistically allowing for the resolution
limits. Also \citet{2015MNRAS.446..521S} have eliminated the boost factor and regulate
the Bondi rate with the ratio of the Bondi to viscous time-scale
\citep{2015MNRAS.454.1038R}. Still, the same limitations apply for the
temperature and density structure of the nuclear gas.

It has been proposed by \citet{1989Natur.338...45S} that black holes
might be primarily fed by gas driven to the center by gravitational
torques from non-axisymmetric perturbations (see also
e.g. \citealp{2013MNRAS.434..606G,2011ApJ...741L..33B,2011MNRAS.415.1027H}). \citet{2011MNRAS.415.1027H}
argue that 
'torque limited' accretion behaves qualitatively different to other
accretion models 
and produces reasonable scaling relations with a smaller
scatter. The parametrized version of this accretion model is a bit
more complicated and less straigh forward to be included in larger
scale cosmological simulations, but the first attempts are promising
(\citealp{2013ApJ...770....5A,2016arXiv160308007A}). Clearly the
choice between the two approaches should be driven by the ratio of the
amnount of angular momentum in the gas to be accreted with the 'torque
limited' model when the ratio of the centrifugal radius to
the Bondi radius becomes larger than unity.   

Approaches following the traditional feedback models assume that some
fraction ($\epsilon_{\mathrm{therm}} \sim 0.05$) of the bolometric luminosity
$L_{\mathrm{bol}} = \epsilon_r dM_{\mathrm{BH}}/dt c^2$
\citep{1973A&A....24..337S,1982MNRAS.200..115S}, with c 
being the speed of light and a radiative efficiency of $\epsilon_r =
0.1$ \citep{1973A&A....24..337S}, is converted into and deposited as
thermal energy in the surrounding ISM  such that the energy
injection rate is $dE_{\mathrm{therm}}/dt = \epsilon_{\mathrm{therm}} \epsilon_{r}
\dot{M}_{\mathrm{BH}} c^2$. \citet{2007MNRAS.380..877S} proposed a
'jet bubble' modification to this simple model depending on the gas accretion
rate onto the black hole. For accretion rates above 1\% of the
Eddington rate the usual fraction of $\epsilon_{\mathrm{therm}} \epsilon_{r}$ of the
accreted rest mass energy is deposited as thermal energy to the
surrounding medium. For lower accretion rates the feedback efficiency
is increased from 0.5\% to 2\% of the rest mass energy and is injected
into heated off-center bubbles that can then buoyantly rise (see
\citet{2010MNRAS.401.1670F}, \citet{2014MNRAS.442.2304H}, and
\citet{2014MNRAS.444.1518V} for slightly modified versions). Such
models are designed to mimic the observed jet induced bubble formation
\citep{2006MNRAS.366..417F,2012ARA&A..50..455F} and, due to the
enhanced coupling efficiency at low accretion rates, i.e. low 
black hole growth rates (see
e.g. \citealp{2005MNRAS.363L..91C,2008MNRAS.388.1011M}), it helps to 
prevent the formation of cooling flows and nuclear star formation in
massive halos
\citep{2007MNRAS.380..877S}. \citet{2012MNRAS.420.2662D}, in a RAMSES
implementation, also identify a 'radio mode' at low accretion
rates and inject kinetic energy into a jet-like bipolar outflow with a
velocity of 10,000 $km s^{-1}$ (see also \citet{2004MNRAS.348.1105O}
and \citet{2006ApJ...643..120B} for the impact of jet feedback in
isolated models).    

Originally developed for SPH, the \citet{2007MNRAS.380..877S} AGN
feedback model has also been used in recent large scale simulations
with the moving mesh code AREPO extended by the influence of the
radiation field of the accreting black holes on the cooling rate of
the gas \citep{2015MNRAS.452..575S,2014MNRAS.444.1518V}. Here the
effect of AGN feedback on reducing the galactic stellar masses was
confirmed, however, at the cost of depleting the massive halos of gas
due to the 'jet bubble' feedback, inconsistent with observations
\citep{2014MNRAS.445..175G}. In general, these relatively straight
forward models not only give reasonable galaxy and black hole 
masses at the high mass end, they also result in plausible evolutions
of the black hole populations and AGN luminosity functions across 
cosmic time
\citep{2013MNRAS.428.2966P,2014MNRAS.442.2304H,2015MNRAS.452..575S,2015MNRAS.450.1349K}.     

Instead of using a constant boost factor, \citet{2009MNRAS.398...53B}
scale $\alpha_{\mathrm{boost}}$ with the  local density for high
ambient gas densities and
set it to unity for low densities - in combination with a stiff
equation of state for dense gas (\citealp{2008MNRAS.383.1210S}, see
\citep{2009MNRAS.398...53B} for a detailed disucssion). The
thermal feedback  is regulated such that the black hole stores the
energy until the 
surrounding particles can be heated to a certain high temperature (in
this case $\gtrsim 10^8$ K), disfavoring rapid cooling of the gas and making
the immediate feedback efficient by construction. The approach is
similar to the stellar 'stochastic thermal' feedback discussed in Section
\ref{stellarfeedback}. It is clear that the density scaling of 
$\alpha_{\mathrm{boost}}$ makes the accretion more sensitive to the 
feedback, which in turn strongly affects the density. In general,
however, the introduction of a temperature limit for the black hole
feedback results in a stronger effect than for the
\citet{2005MNRAS.361..776S} model with no 
need for changing the energy conversion efficiency and the feedback
methodology. This model also results in reasonable baryon fractions and
stellar masses for massive galaxies and 
black hole masses \citep{2015MNRAS.446..521S}. In particular it
reproduces reasonable gas fractions and X-ray luminosities for galaxy
groups
\citep{2010MNRAS.406..822M,2014MNRAS.441.1270L,2014MNRAS.442..440C}
which are over-predicted by models with a constant boost factor
and thermal feedback.   

An Eddington limited thermal black hole feedback scheme similar to
\citet{2009MNRAS.398...53B} with a density dependent
$\alpha_{\mathrm{boost}}$ and the same underlying equation of state
for the gas, has also been implemented in the adaptive mesh refinement code
RAMSES \citep{2011MNRAS.414..195T,2012MNRAS.420.2662D}.  The black holes are not seeded in
halos above a certain mass but the accreting sink particles are
generated when certain conditions on the stellar component and gas
density are met (\citealp{2011MNRAS.414..195T}, see
e.g. \citealp{2012MNRAS.420.2662D,2012ApJ...745L..29D,2013MNRAS.428.2966P,2014MNRAS.442.2304H}
for different implementations of black hole seeding). 

Zoom simulations of group and cluster sized halos in particular have highlighted
the impact of AGN feedback on reducing the stellar mass in the central
galaxies by preventing cooling flows and subsequent star formation 
\citep{2007MNRAS.380..877S,2010ApJ...720L.149T,2010MNRAS.406..822M}. AGN
feedback also has important consequences for the hot gas in these
halos. By removing low entropy gas at higher redshifts (at the peak of
the black hole growth) AGN bring the simulated present day hot gas
properties in much better agreement with observations of thermal X-ray
emission \citep{2010MNRAS.406..822M,2011MNRAS.412.1965M,2014MNRAS.441.1270L}. 

The above simulations also indicate a potentially interesting effect of
AGN feedback on the stellar kinematics. As the amount of gas cooling
and subsequent star formation in the central galaxies is efficiently
suppressed \citep{2012MNRAS.420.2859M,2012MNRAS.422.3081M}, the ratio
of in-situ formed to accreted stars is significantly reduced, 
increasing the size of the system but also reducing the amount
of rotational support
\citep{2013arXiv1301.3092D,2014MNRAS.443.1500M}. AGN feedback
transforms BCGs from fast rotators to  slow rotators, possibly in
better agreement with observations
\citep{2013MNRAS.433.3297D,2014MNRAS.443.1500M}. This highlights the
potentially important role of AGN feedback for regulating the ratio of
in-situ star formation to accretion
\citep{2012MNRAS.419.3200H,2013MNRAS.433.3297D} determining the
stellar kinematics of the systems and regulating the stellar density
distributions.    

The strength of these kinematic signatures will not only be influenced by
feedback from AGN and stars, but also by the mass of the galaxies 
and their environment. Low mass, star forming disk galaxies favored in
low-density environments, predominantly grow by accretion of gas and
subsequent in-situ star formation, and are affected by stellar
feedback, and less so by AGN. Higher mass, early-type galaxies form in
high-density environments - primarily affected by feedback from accreting
super-massive black holes - and their late assembly involves merging
with other galaxies, which might also be of an early type
(e.g. \citealp{2016MNRAS.458.2371R,2017MNRAS.464.1659Q}). So far it 
has not been possible to perform a statistically meaningful comparison
of kinematic properties of galaxy populations to observed population
properties, like the observed increasing fraction of slow versus fast
rotators for early-type galaxies as a function of environmental
density \citep{2011MNRAS.416.1680C}. With the newly performed Eagle
\citep{2015MNRAS.446..521S} and
Illustris \citep{2014MNRAS.444.1518V} simulations this
might now be possible for the first time, due to the large simulated
volume and the relatively high spatial resolution attained ($\lesssim
1kpc$). Not only can the simulated two-dimensional kinematic
properties be compared to observations, it will also be possible to
make (statistical) connections to characteristic formation histories,
properties of progenitor galaxy populations and to investigate trends
with environment. This will not only be limited to stellar kinematics
but will also include gas kinematics and metallicity and
two-dimensional stellar abundance patterns. In addition, we will be
able to assess the impact of the major feedback mechanisms (from
massive stars and AGN) on the kinematic properties of high and low
mass galaxies in all environments, a study that will be supported by
higher resolution, cosmological zoom simulations for characteristic
cases. Also several recent papers
(e.g. \citealp{2010MNRAS.407.1403C,2013ApJ...772...97B,2014MNRAS.441.1270L,2014MNRAS.442..440C,2014MNRAS.441.1270L,2015MNRAS.451.3868L}) 
have shown that a proper prediction of the thermal X-ray halos of
massive galaxies provides a stringent test of the correctness of any
implemented feedback scheme. We discuss mechanical and radiative AGN
feedback models in Sec. \ref{mech}.  

\section{The need for accurate modelling of the galactic 
  interstellar medium and 'feedback'}
\label{ISM}

\begin{figure}
  \centering
  \includegraphics[width=1\linewidth]{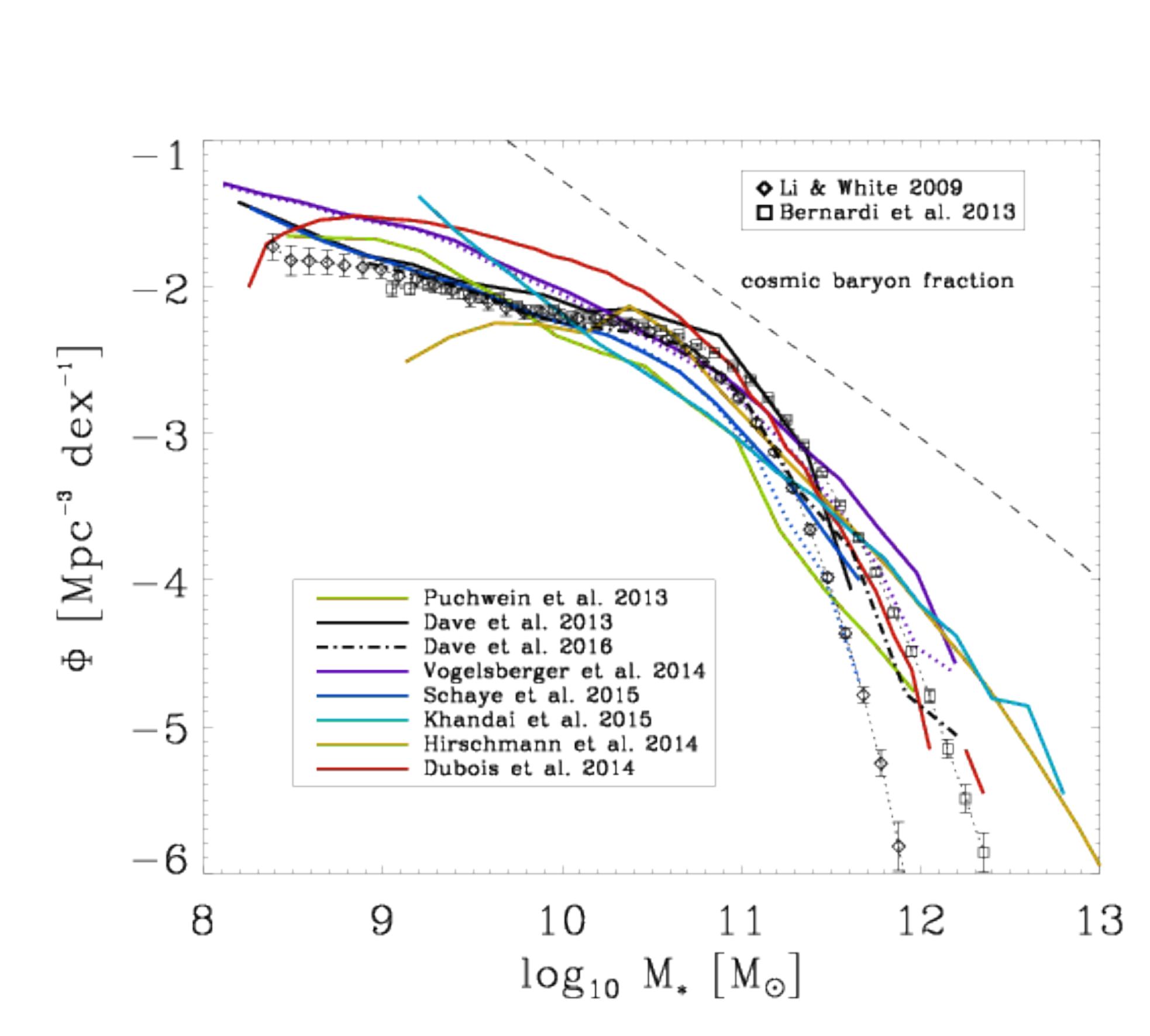}

\caption{Comparison of galaxy stellar mass functions from recent large
  scale cosmological simulations of representative volumes of the 
  Universe (see text). The simulations include stellar and AGN
  feedback with the exception of \citep{2013MNRAS.434.2645D} who use an
  empirical heating model in massive halos. The different groups
  typically adjust the key parameters in the varying sub-resolution
  models to match observations of galaxy mass functions like the one of
  \citet{2009MNRAS.398.2177L}. For reference we show an alternative
  mass function with different mass estimates for massive galaxies
  \citep{2013MNRAS.436..697B}. At a given 
  mass the abundance can vary by up to an order of magnitude,  still
  considering the range in spatial resolution (from 0.5 kpc to 3 kpc)
  and the significant difference in sub-resolution  models the
  agreement between the simulations is remarkable for some models. The
  dashed line for  \citet{2014MNRAS.444.1518V} and
  \citet{2015MNRAS.446..521S}  indicate different mass estimates. The
  dashed line shows the hypothetical galaxy mass function assuming the
cosmic baryon fraction.}   
\label{mass_function}
\end{figure}

The large differences of sub-resolution models presented in the previous
section are noteworthy but also a bit worrying as the models should be
representations of the same underlying physical processes. Many groups
have been able to design and callibrate feedback procedures that allow them to more
or less successfully match the present day galaxy mass function, or
more accurately, the ratio of dark matter to galaxy masses, the
cosmological evolution of galaxy abundances, and the evolution in
galaxy sizes and stellar populations (the Eagle project seems to have
done this most successfully). In
Fig. \ref{mass_function} you can assess to which level the different
groups have succeeded in this calibration to present day galaxy mass
functions similar to the one presented in
\citet{2009MNRAS.398.2177L}. With the large 
conceptual differences in the respective feedback models the good (or
not so good) agreement with the observed mass function can be  
attributed to a more or less successful tuning of the 
normalizations and scalings in the sub-resolution models. The
theoretical predictions are stellar mass functions from large scale
cosmological simulations in recent publications: from
\citet{2013MNRAS.434.2645D} using  
traditional SPH, 'energy and momentum driven' decoupled winds and a
heuristic model for gas heating in massive halos (no AGN feedback);
from \citet{2016arXiv160401418D} using a meshless finite mass method
with star formation feedback scaling motivated by higher resolution
simulations and an improved empirical gas heating model to
\citet{2013MNRAS.434.2645D} for massive halos; from
\citet{2013MNRAS.428.2966P} using traditional SPH with momentum driven
winds and thermal and 'bubble' AGN heating; from
\citet{2014MNRAS.444.1453D} (Horizon-AGN) with AMR, mechanical
supernova feedback, thermal AGN feedback with variable   
boost factor and jet-feedback for low accretion rates, from
\citet{2014MNRAS.444.1518V} (the Illustris simulations) with a moving mesh code, 
decoupled 'momentum driven' winds, thermal and 'bubble' AGN heating and
radiation input from the AGN; \citet{2015MNRAS.446..521S} (the Eagle
simulations) with improved SPH, 'stochastic thermal' heating from
stars and AGN; from \citet{2014MNRAS.442.2304H} (the Magneticum simulations)
with traditional SPH, constant winds, thermal AGN heating and a
modified 'bubble' heating; from \citet{2015MNRAS.450.1349K} (the
MassiveBlack II simulations) with traditional SPH, 'decoupled wind'
and thermal AGN feedback. The simulations have more or
less succeeded in this exercise. For comparison we show a (typical) observed
galaxy mass function in the local Universe
\citep{2009MNRAS.398.2177L}. Again, we point out that such a mass
function is used for most simulations and alternative mass functions
take extended stellar mass 
distributions in massive galaxies into account
\citep{2013MNRAS.436..697B}. Even with the significant differences in 
the sub-grid model assumptions it seems that many simulations capture
the basic characteristics. Still to be achieved are cosmological simulations
yielding good matches to galaxy population properties on the basis of
numerically resolved ab initio physical modeling of feedback processes.          

Returning to the outline of the physical problems encountered in
studying galaxy formation and evolution we had already noted earlier
that for Part (B) of the problem - feedback - there were a number of
physical processes that we know are important but remain
unsolved. Primary among them is the questions of which physical
processes regulate the multi-phase structure of the ISM and what is
the main driver for galactic outflows.

\subsection{Supernova explosions} 
\label{supernovae}

Core-collapse supernova explosions have for long been the primary
suspect to play a crucial role in galaxy formation
\citep{1974MNRAS.169..229L,1986ApJ...303...39D,1993MNRAS.265..271N}. During
these singular and final events in a massive star's live typically 2 - 5
$M_{\odot}$ of gas are ejected into the ambient interstellar medium
(ISM) at supersonic velocities of $v_{eject} \sim 6000 - 7000 km
s^{-1}$ \citep{2012ARNPS..62..407J} driving a shock into the ambient
ISM. Apart from the injection of metals, supernovae can - in the
energy conserving phase of the blast wave - 
heat about three orders more ambient mass than their ejecta to high
temperatures. This makes them the prime ources of hot ($T \sim 10^6 K$)
gas in the star forming ISM.  By creating the hot, X-ray emitting,
phase they impact the large-scale multi-phase structure of the ISM 
\citep{1977ApJ...218..148M,2015MNRAS.454..238W,2015ApJ...814....4L} and might be important
for driving galactic outflows, fountain flows, and galactic winds
through hot, low density, chimneys
\citep{1989ApJ...345..372N,1985Natur.317...44C,1990ApJS...74..833H,2000MNRAS.314..511S,2003RMxAC..17...47H,2006ApJ...653.1266J,2012ApJ...750..104H,2014A&A...570A..81H,2016MNRAS.456.3432G}.  
The momentum injected by supernovae contributes to the kinetic energy
content of the ISM. With pure momentum injection simulations it has been
argued that supernovae can create realistic turbulence (see reviews on
ISM turbulence by \citealp{2004ARA&A..42..211E,2004ARA&A..42..275S}) in the warm
and cold ISM and regulate the scale heights of galactic disks (by
large scale turbulent pressure) as well as their star formation rates (see
\citealp{2011ApJ...731...41O,2015ApJ...815...67K}). 

The importance of supernova explosions for setting the ISM
structure motivates a more detailed review of the different phases of
supernova blast waves (see also
\citealp{1982ApJ...258..790C,1988RvMP...60....1O,1988ApJ...334..252C,1998ApJ...500..342B,2011piim.book.....D,2015ApJ...802...99K,2016MNRAS.460.2962H}). The 
direct momentum of supernova ejecta is insufficient to accelerate a
significant amount of gas to high velocities in the early {\it free 
expansion phase}. Once the supernova ejecta have swept up cold interstellar
material of comparable mass the remnant enters the energy conserving
{\it Sedov-Taylor phase}
\citep{1959sdmm.book.....S,1950RSPSA.201..159T,1999ApJS..120..299T}. 
In this phase about 1000 times the ejecta mass is heated and
about 10 times the initial (ejecta) radial momentum can be generated
as long as the temperature changes are dominated by adiabatic
expansion (this can amount to 100 times the ejecta momentum in the
absence of cooling). As soon as radiative
losses become dominant a cooling shell forms and the amount of hot gas 
decreases rapidly. Analytical estimates for the time
$t_{\mathrm{sf}}$, radius $r_{\mathrm{sf}}$, velocity
$v_{\mathrm{sf}}$,
temperature $T_{\mathrm{sf}}$, mass $M_{\mathrm{sf}}$, and radial
momentum $p_{\mathrm{sf}}$ at shell formationas a function of
explosion  
energy $E_{\mathrm{51}}$ in units of $10^{\mathrm{51}} ergs$, and
the number density of a homogenous ambient medium in $cm^{-3}$ result
in the following relations (taken from \citet{2015ApJ...802...99K},
but see also \citealp{2011piim.book.....D}): 

\begin{eqnarray}
  t_{\mathrm{sf}} & = & 4.4\times 10^4 yr E_{51}^{0.22} n_0^{-0.55} \label{tsf}\\
  r_{\mathrm{sf}} & =  & 22.6pc\, E_{51}^{0.29}\,n_0^{-0.42} \label{rsf} \\
  v_{\mathrm{sf}} & =  & 202km s^{-1}\,  E_{51}^{0.07}\, n_0^{0.13}\label{vsf} \\
  T_{\mathrm{sf}} & =  & 5.67 \times 10^5 K\,E_{51}^{0.13}\,n_0^{0.26} \\
  M_{\mathrm{sf}} & =  & 1680M_{\odot}\, E_{51}^{0.13} \,n_0^{-0.26} \label{msf}\\
  p_{\mathrm{sf}} & =  & 2.17 \times 10^5 M_{\odot} km s^{-1} \, E_{51}^{0.93}\,n_0^{-0.13} \label{psf}
\end{eqnarray}

\citet{2015ApJ...802...99K} have shown in detail that these analytic
estimates agree well with direct, high-resolution, three-dimensional
numerical simulations (see also
\citet{2015MNRAS.451.2757W,2015A&A...576A..95I,2015MNRAS.450..504M}).    

After shell formation the supernova enters a short {\it transition
  phase} and the following {\it pressure driven snowplow phase} is
powered by the homogenous pressure inside the shell. Once all excess
thermal energy is radiated away, no radial momentum can be generated
any more and the remnant enters the {\it momentum conserving snowplow
phase}. Travelling into the interstellar medium the shock wave
transforms into a sound wave (when the expansion velocity reaches the
sound speed of the interstellar medium) and fades away. It can be
shown that for solar neighborhood conditions an initially uniform
medium will be completely changed within 2Myrs by overlapping remnants
in their fade-away stage \citep{2011piim.book.....D}. This simple
argument indicates that supernovae alone might determine the thermal
and dynamical state of the ISM \citep{1977ApJ...218..148M}.
  Subsequent to this phase supernova remnants will propagate in the
  multi-phase medium with greater efficiency and reduced losses. These
  properties have yet to be fully assimilated into cosmological
  simulations of galaxy formation. 

Radiative cooling, i.e. the actual ambient density and metallicity at
the time of the supernova explosions, determines the duration of the
momentum generating phases of which the {\it Sedov-Taylor phase} is
the most important. For single supernovae exploding in ambient
densities of $\sim 100 - 0.1 cm^{-3}$ cooling becomes dominant after
about $\sim 10^{4} - 10^{5.5}$ years limiting the momentum  generation
to factors of $\sim 10 - 30$ \citep{2016MNRAS.460.2962H}. For reliable
simulations of the galactic ISM it is important that the momentum
generating phases of supernovae remnants can be captured
accuratly. While analytic estimates are useful for homogeneous ambient
media they cannot simply be applied to the more complex multi-phase
structure of a realistic ISM. Here, numerical simulations have made a
significant progress in recent years. We discuss this effort a bit
more in detail as we think it is a good example of how resolved numerical
simulations with different simulation codes can be used to understand
a specific relevant physical process in more complex environments. 

\begin{figure}
  \centering
  \includegraphics[width=1\linewidth]{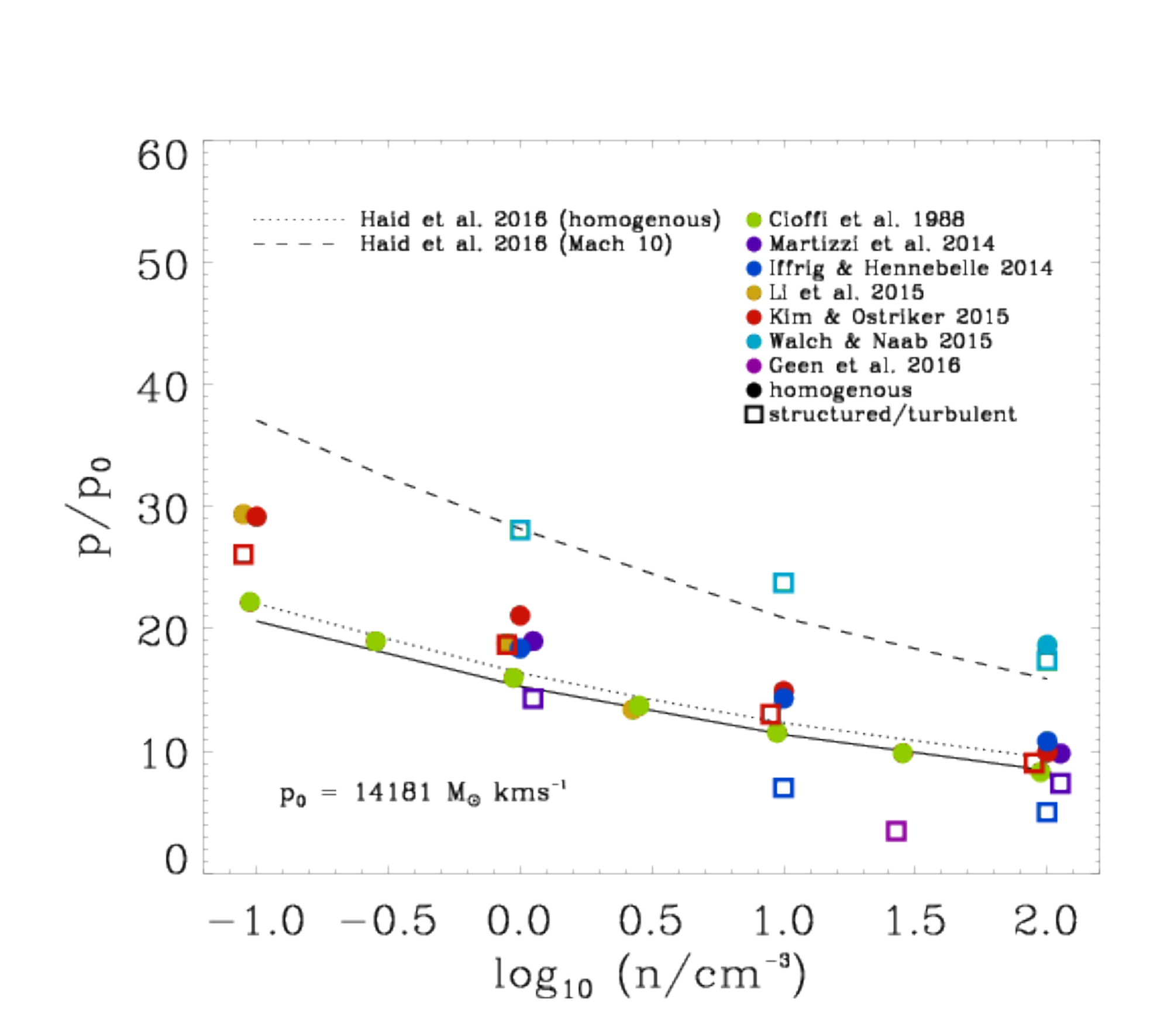}
\caption{Momentum generated in radiative supernova remnants for various ambient
  densities normalized by a fiducial initial momentum of $p_0 = 14181
  M_{\odot} km s^{-1}$ for an explosion energy of $10^{51} erg$ and
  two solar masses ejecta. The analytically derived momentum at shell
  formation (Eqn. \ref{psf}, solid line) terminates the energy
  conserving Sedov-Taylor phase \citep{2015ApJ...802...99K}  and the
  momentum can increase a bit more until the beginning of the momentum
  conserving snowplow phase (dotted line). The dashed line
  indicates the momentum injection for an analytical model of a
  log-normal (Mach 10) density distribution
  \citep{2016MNRAS.460.2962H}. Colored symbols show results from three
  dimensional numerical simulations (with the 
  exception of the one-dimensional simulations by
  \citealp{1988ApJ...334..252C}) with homogenous, structured or
  turbulent ambient media carried out with three different grid codes
  and a particle based SPH code.}
\label{sn-momentum}
\end{figure}

In Fig. \ref{sn-momentum} we show an
overview of mostly three-dimensional numerical simulations measuring
the momentum injection into the interstellar medium for the various
conditions of the ambient ISM. \citet{2015MNRAS.450..504M} have used
the adaptive mesh refinement code \textit{RAMSES} for homogeneous
ambient medium and one with a lognormal density distribution
representing a Mach 30 turbulent ISM. The simulations of
\citet{2015ApJ...815...67K} have been performed with the
\textit{ATHENA} code for a homogenous and a structured two-phase
medium (cold and warm phase). Additional simulations for  a
three-phase medium have been performed by \citet{2015ApJ...814....4L}
with the AMR code \textit{ENZO}. We have to note that even at low
resolution the total mometum input of a supernova can be correctly
computed. However, the swept up mass and the velocity of the shell can
still be incorrect \citep{2016MNRAS.458.3528H}    

A supernova does not only inject momentum into the ISM. It also
generates hot gas in the early phases of the remnant's evolution. The
maximum amount of hot gas is reached at the time of shell formation,
marking the end of the Sedov-Taylor phase. If no other supernova
explodes within the remnant's radius until the time of shell formation
$T_{\mathrm{sf}}$ the remnant will cool rapidly and no stable hot
phase can be generated. For a homogenous ISM of density $n_0$ and a
given supernova rate density $S$ we can estimate the expectation
value $N_{\mathrm{hot}}$ for a supernova to explode in a hot phase within the shell
formation radius $r_{\mathrm{sf}}$: 

\begin{eqnarray}
  N_{\mathrm{hot}} & = & S \frac{4\pi}{3} r_{sf}^3 t_{sf}. 
\end{eqnarray}

With Eqns. \ref{rsf} and \ref{msf} this results in

\begin{eqnarray}
  N_{\mathrm{hot}} & = & S 2.13 \times 10^{-6} kpc^3 Myr E_{51}^{1.09} n_0^{-1.81}. \label{fhot}
\end{eqnarray}

For a typical gas surface density similar to the solar neighborhood of
$10 M_{\odot} pc^{-2}$ the Kennicutt relation
\citep{1998ApJ...498..541K} gives an observed star formation 
rate surface density of $ 7 \times 10^{-2} M_{\odot} yr^{-1}
kpc^{-2}$. With a Salpeter like initial mass function and an assumed 
  disk thickness of $250 pc$ the resulting supernova rate density is
$S\sim 280 kpc^{-3} Myr^{-1}$. For an average gas number density of
  $n_0 = 1 cm^{-3}$ the expectation value would be only
  $N_{\mathrm{hot}} \sim 6 \times 10^{-4}$. Due to the strong power-law dependence on the
  density in Eqn. \ref{fhot} a lower density of $n =
  0.015   cm^{-3}$ results in $N_{\mathrm{hot}} \gtrsim 1$ and a stable
  volume filling hot phase can form. Once such
  a condition is reached and the system 
cannot vent the hot gas it will undergo a {\it thermal
  runaway}. Subsequent supernovae explode in even hotter and lower
density regions with even less thermal losses and larger shell
formation radii. Once the volume is dominated
by over-pressured hot gas no shell will form, cooling losses are
minimal and most of the mass is in small cold clumps. This process has
recently been described by \citet{2015MNRAS.449.1057G} and 
\citet{2015ApJ...814....4L} with hydrodynamical simulations in
periodic boxes. If the ISM can vent the hot gas, an outflow is
driven. 

\begin{figure}

  \centering
  \includegraphics[width=1\linewidth]{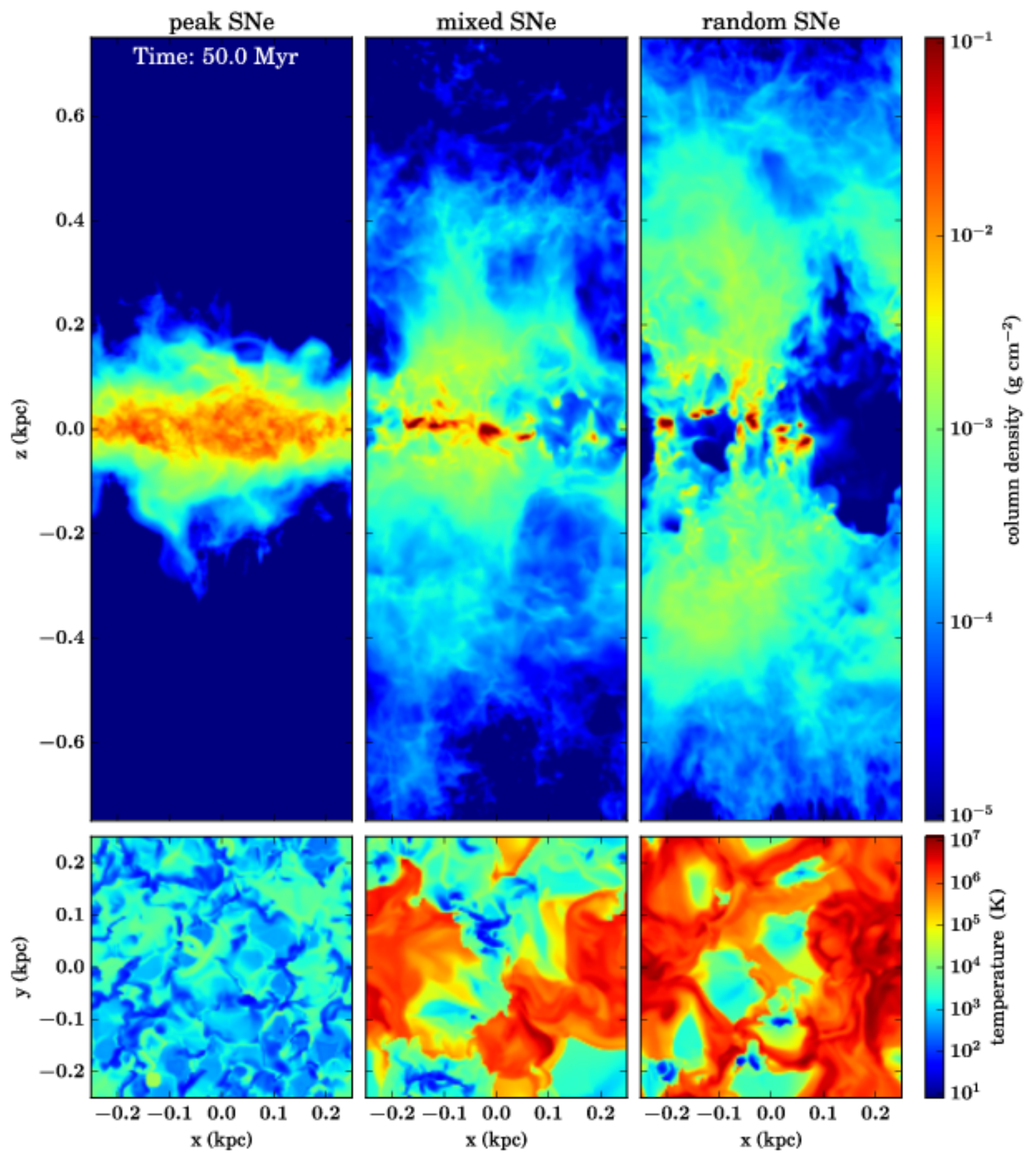}

\caption{Snapshots of the vertical gas column density distribution (top
  panels) and mid-plane temperatures (bottom panels) for three
  simulations of stratified galacitc disk ($ \Sigma_{\mathrm{gas}} =
  10M_{\odot} pc^{-2}$ shaped by supernovae (SNe) exploding at a constant
  rate (taken from the SILCC simulations,
  \citealp{2015MNRAS.454..238W,2016MNRAS.456.3432G}). In peakSNe (left 
  panels) each supernova explodes at the current density peak. Rapid
  thermal losses limit the SN impact to momentum ejection driving an
  (unrealistic) turbulent two-phase medium with no outflows 
  and no hot gas. With 50 per cent of the supernovae exploding at random
  positions at lower densities a hot phase appears (mixed SNe, middle
  panel). If all SNe explode at random positions the hot phase becomes
  volume filling and drives a vertical outflow (random SNe, right
  panles, see \citealp{2016MNRAS.456.3432G}). This figure clearly
  illustrates the strong impact of the actual location of SN
  explosions on the multi-phase structure of the ISM and the driving
  of outflows (see Eqn. \ref{fhot}). We note that non of these
  variations can be captured in current large scale cosmological
  simulations which have single resolution elements of the size of the
  above simulation boxes ($\sim 500 pc$) and rely on sub-resolution models.}       
\label{silcc}
\end{figure}

The strong dependence of the hot gas volume filling fraction in
Eqn. \ref{fhot} on the environmental density of the supernova
explosions has significant consequences for the evolution of the 
galactic ISM. In Fig. \ref{silcc} we illustrate this with three numerical
experiments (part of the SImulating the Life Cycle of
molecular Clouds [SILCC] simulation project, \citealp{2015MNRAS.454..238W}). The setup is a
stratified galactic disk with a surface density of
$\Sigma_{\mathrm{gas}} = 10M_{\odot} pc^{-2}$  embedded in a stellar 
disk potential. The initially homogenous ISM is driven by supernova
explosions at a constant rate based on observational estimates. If all
supernovae explode at the current density peaks (typical densities of
$n = 100 cm^{-3}$, see \citealp{2016MNRAS.456.3432G}) the explosions
suffer from radiative losses, no hot phase develops and the impact is
limited to momentum injection. Rapidly  a turbulent, pressure
supported two phase (warm and cold gas) medium develops and the scale 
height is set by the turbulent pressure. \citet{2011piim.book.....D} presents 
a simple estimate for this process to take only $2 Myrs$. This behavior is reported
from pure momentum injection models
\citep{2011ApJ...731...41O,2011ApJ...743...25K,2013ApJ...776....1K,2015ApJ...815...67K}. The
major shortcomings of these models are that no hot phase can develop
and the cold phase cannot become dense enough for molecular gas
formation (see e.g. \citealp{2015MNRAS.454..238W}). If some half of
the supernovae do not explode in density peaks but rather at random
locations in the disk, as would be expected from the number of
early-type 'runaway' stars discussed below, the ambient density
distribution for supernovae becomes
bimodal. Random supernovae in low density regions $n \sim 0.1 cm^{-3}$
compress the cold gas to higher densities $n\gtrsim 100 cm^{-3}$, where the peak
supernovae explode. The system can form a hot phase (see
middle panels of Fig. \ref{silcc}). Once all supernovae explode at
random positions (most ambient densities $n \lesssim 0.1 cm^{-3}$) the ISM
becomes highly structured and rapidly develops a stable hot phase,
which is  expanding into the halo and drives a galactic outflow (see
\citet{2016MNRAS.456.3432G} for the models shown 
in Fig. \ref{silcc}). This behavior has already been
reported by pioneering three-dimensional hydrodynamical simulations of
stratified disks by 
\citet{1999ApJ...514L..99K}, \citet{2000MNRAS.315..479D},
\citet{2004A&A...425..899D}, and \citet{2006ApJ...653.1266J}. The
major shortcomings of these type of models are that they neglect the
galactic environment (radial gas flows and inflows) and, due to the
idealized gemoetry, gas flows into and out of galactic halos cannot be
modelled accurately (see e.g. \citealp{2016MNRAS.459.2311M}).

OH maser measurements indicated that only about 15 per
cent of core collapse supernovae interact with dense molecular
gas \citep{2009ApJ...706L.270H}. Therefore the typical ambient density
for explosions is lower than the dense birth places of massive
stars. Now the questions is which astrophysical processes determine
the ambient densities of supernova explosions? There are two phenomena
which can result in this: the massive stars move away from their dense birth places into
gas with lower volume densities and larger volume filling fractions
and/or the stars change their environmental densities by stellar
winds, ionising radiation and clustered supernova explosions.  

Most, if not all, stars form in star clusters and are expected to be
temporally and spatially correlated, eventually driving super-bubbles
with more efficient energy coupling and momentum generation 
(e.g. \citealp{1988ApJ...324..776M,2006ApJ...653.1266J,2014MNRAS.443.3463S}).
The typical velocity dispersion in newly born star 
clusters is $\sim 1 km s^{-1}$ or $1 pc Myr^{-1}$. Assuming the
cluster becomes unbound  massive stars can travel up to $40 pc$ before
they explode (assuming typical supernova delay times for single stellar
populations). These 'walkaway' stars can therefore easily leave their
dense birthplaces and explode in lower density regions.  As
most of the volume of the ISM is not in the cold phase most supernovae
might explode in the warm or hot phase, significantly changing the coupling
efficiency \citep{2005MNRAS.356..737S,2009ApJ...695..292C}. There is also
the 'runaway' star population \citep{1986ApJS...61..419G} (about 45 \%
of O-stars and 15 \% of B-stars are 'runaways') with typical
velocities of $\sim 30 - 40 km s^{-1}$ and maximum velocities as high
as a few hundred $km s^{-1}$
(e.g. \citealp{2011MNRAS.411.2596S}). These high-velocity stars 
originate from close binary 
systems becoming unbound by a supernova explosion \citep{1957moas.book.....Z,1961BAN....15..265B} and/or from
dynamical interactions in dense regions of star clusters \citep{1967BOTT....4...86P}. They can
travel up to several hundreds of parsecs away from their birthplaces far into
inter-arm regions or the lower galactic halo. Their explosion locations
in galactic disks can therefore be considered as almost random, similar to
the explosions of SNe Ia which contribute about 20 -25 \% to the supernova
rate in the solar neighborhood \citep{1994ApJS...92..487T}. They
explode independently of the gas mass distribution within the ISM, a
process approximately taken into account in high-resolution simulations of  
stratified disks which are most useful to study the launching of
outflows and, at the same time, create a realistically structured ISM
(e.g. \citealp{2005A&A...436..585D,2006ApJ...653.1266J,2012ApJ...750..104H,2016arXiv161008971L}).  
Such detailed small scale simulations with a realistic ISM structure
will help to bridge the gap in scale and physical understanding  to
galaxy scale simulations.   

\subsection{Stellar winds}

Massive stars themselves also impact their ambient medium. Radiation
driven stellar winds from O- and B-stars 
\citep{1975ApJ...195..157C,1996A&A...305..171P,2000ARA&A..38..613K}
create bubbles of low density gas around the stars. Typical B-stars
with masses $\sim 9 M_{\odot}$, mass-loss rates of
$\dot{M}_{\mathrm{wind}}  \sim 10^{-9} M_{\odot} y^{-1}$, and wind velocities
  of $v_{\mathrm{wind}} \sim 2000 km s^{-1}$ have an intergrated wind
luminosity of only a few times $\sim  10^{47} erg$. However, very massive
stars can reach as much as $E_{\mathrm{wind}} \sim 10^{51}
erg$. Although energetically much less important than supernova
explosions, stellar wind blown bubbles can significantly reduce the gas
densities around massive stars and thereby increase the impact of the
supernovae. Furthermore, since momentum injection goes as $\dot{E}/v$,
winds from massive stars can contribute more direct momentum than the
supernovae themselves. Interestingly, stellar winds can also significantly reduce
the star formation process in forming star clusters
(\citealp{2008MNRAS.391....2D}, although
\citet{2013MNRAS.436.3430D,2014MNRAS.442..694D} argue that the
combined effect of stellar winds and ionising radiation only has 
modest impact on star formation) and might be a stronger regulator for
galactic star formation than supernovae \citep{2016arXiv160605346G}.    

\subsection{Radiation}

Stellar evolution models indicate that the total energy released by
newly formed stellar populations is, by a large margin of two orders of
magnitude, dominated by stellar radiation, which itself is dominated
by massive stars (e.g. \citealp{1999ApJS..123....3L}). 
By the time the first supernova, with a canonical energy of $10^{51}
erg$, has exploded, the stars would have already emitted $\sim 10^{53}
erg$ as radiation, and $\gtrsim 10^{50} erg$ in stellar winds. It has
been realized lately that, in galaxy scale simulations, 
accounting for the stellar luminosity (and winds) might significantly
enhance the coupling of the stellar energy and momentum output to
the ISM
(e.g. \citealp{2011MNRAS.417..950H,2012MNRAS.421.3522H,2013ApJ...770...25A,2014MNRAS.444.2837R}). 
The generation of momentum is of particular interest as it cannot
easily be radiated away like thermal energy. The efficient cooling results in
only a moderate, 10\%, contribution of the thermal energy density in the ISM
\citep{1990ApJ...365..544B}. 
Due to the generally severe radiation losses in dense interstellar
environments it is, however, unclear, how much of the injected  energy
can be converted into momentum. If this process is efficient, then
stellar radiation might significantly contribute to the driving of
turbulence and the launching of
galactic winds
(e.g. \citealp{2005ApJ...618..569M,2010ApJ...709..191M,2011ApJ...735...66M,2015MNRAS.448.3248G,2015ApJ...804...18A}).   

Ionizing UV photons create HII regions around young massive stars by
heating the parental cloud from $\lesssim 100 K$ to $\sim 10^{4}K$. At
this stage the dynamics of the ISM is dominated by the thermal
pressure of the ionized gas with sound speeds $\lesssim 10km
s^{-1}$. The momentum input by direct absorption of UV photons (single
scattering) seems sub-dominant
\citep{1969ApJ...157..583M,1978ppim.book.....S,2004ApJ...608..282A,2009ApJ...703.1352K,2014MNRAS.439.2990S}. 
It might be sufficient to drive turbulence at a low level
\citep{2009ApJ...694L..26G} and even disrupt small clouds
on short time-scales \citep{2006ApJ...653..361K,2012MNRAS.427..625W},
however, over-dense regions of the surrounding ISM  will also be
compressed into clumps and pillars
\citep{2007MNRAS.375.1291D,2010ApJ...723..971G,2012MNRAS.427..625W}
making further coupling more difficult.        
    
A full radiation transfer treatment of ionizing radiation from
massive stars in galaxy scale simulations is technically challenging
(see e.g. \citealp{2012MNRAS.427..311W} for single scattering). 
It has been approximated in a Str\"omgren approach, i.e. the ISM 
around massive stars corresponding to a Str\"omgren sphere
\citep{1939ApJ....89..526S} is ionized and heated to $\sim 10^{4} K$
(e.g. \citealp{2013MNRAS.436.1836R,2012MNRAS.421.3522H}) or 
in the optically thin regime (assuming escape fractions from 'clouds')
with radiation field attenuation to follow the impact on gas cooling
(e.g. \citealp{2014MNRAS.437.2882K}). \citet{2014MNRAS.437.2882K} claim
a significant impact of the local UV radiation resulting in a
suppression of star formation for Milky Way like galaxies ($\sim 40
\%$) by increasing the cooling time and the equilibrium temperature. 

It has been argued that the radiation pressure on dust of re-emitted
and scattered infrared radiation can result in a significant momentum
input into the ISM, 
\begin{eqnarray}
\dot{P}_{rad} \sim (1 + \tau_{\mathrm{IR}}) L/c.
\label{prad}
\end{eqnarray}
The efficiency of this process depends on the optical depth to
  the re-radiated long-wavelength emission of the dust,
  $\tau_{\mathrm{IR}}$, i.e. on the details of 
multiple scatterings in optically thick regions surrounding the young
stars. Based on small scale simulations, \citet{2013MNRAS.434.2329K}
argue that the momentum input does not exceed $L/c$ due to the
structure, and therefore inefficient trapping, developing in the ISM
\citep{2012ApJ...760..155K}. However, using a different radiation
transport method, \citet{2014arXiv1403.1874D} find 
a slightly stronger coupling of the radiation. In an attempt to
approximately include this effect in high resolution galaxy scale simulations 
\citet{2011MNRAS.417..950H,2012MNRAS.421.3488H,2012MNRAS.421.3522H}
add momentum to  the surrounding gas either in a stochastic or
continuous way and indicate that they can use this process to drive galaxy scale
winds. \citet{2011MNRAS.417..950H} give the gas particles
initial kicks of the order the escape velocity from local 'gas clumps'
or 'star clusters' (between $60 km s^{-1}$ and $\sim 350 km s^{-1}$ for
'clusters' with masses $\sim 10^5 - 10^9 M_{\odot}$) guaranteeing that
dense regions become unbound before additional radiation pressure and
supernovae can act. Locally they approximately compute $\tau_{\mathrm{IR}}$
from the local gas properties assuming a high dust opacity of $\sim 5
cm^2 g^-1$ (see also \citealp{2014MNRAS.444.2837R}) and use a model
for attenuated radiation to compute the momentum input from all stars
at large distances \citep{2011MNRAS.417..950H}. This empirical
implementation of radiation pressure may result in large scale 
galactic winds by a non-linear interaction of the different feedback mechanisms
with the wind mass-loading changing as a function of galaxy properties
\citep{2012MNRAS.421.3522H}. 

Other groups have followed similar paths to
account for the full energy input of stellar populations and approximate
the effect of radiation pressure. \citet{2013ApJ...770...25A} have
implemented the local momentum input as in Eq. \ref{prad} with photon
trapping acting at early ($t \lesssim 3Myr$) embedded stages.
\citet{2013MNRAS.434.3142A} assume a large, fixed, value for the
optical depth ($\tau_{IR} \sim 25$) and scale the momentum input with
the local gas velocity dispersion and metallicity. In a cosmological context
it has been demonstrated, using different codes, that such efficient 
momentum input and the resulting winds can promote the formation
of disk galaxies with appropriately low conversion efficiencies 
(\citealp{2014MNRAS.445..581H,2015ApJ...804...18A}, see Fig. \ref{disks}).
However, \citet{2014MNRAS.444.2837R} implemented an approximate
radiation transfer for ultraviolet and infrared radiation, where the dust
opacity becomes an important factor to regulate the feedback
efficiency. They argue that the momentum input required to drive
strong outflows at the same time disturbs the gas and the resulting
stellar disk so much that it becomes impossible to retain the flat
disk morphology.  In summary, while many promising calculations have
been made (see e.g. \citealp{2015MNRAS.451...34R} for a first
application of radiative transfer processes in galaxy-scale
simulations), it is not yet clear how much radiation from young stars can 
really contribute to the driving of strong galactic outflows. For a
complete understanding of this process more accurate models for dust
evolution in galaxy formation simulations have to be
considered. Good steps forward have been recently made also in this direction
(e.g. \citealp{2015MNRAS.449.1625B,2016MNRAS.457.3775M,2016ApJ...831..147Z})

\subsection{Magnetic fields and cosmic rays}

Magnetic fields and cosmic rays (CRs), relativistic high-energy
particles, mostly protons and electrons, are an integral, non-thermal
component of the interstellar medium. In the solar neighborhood the
energy density in CRs, magnetic fields and the kinetic energy density are
comparable and significantly higher than the thermal pressure \citep{1990ApJ...365..544B,2001RvMP...73.1031F}.
CR spectra have been measured over many orders of magnitude from
$E_\mathrm{CR}\sim10^{7}\,eV$ up to the energies of
$E_\mathrm{CR}\sim10^{20} eV$. As the galactic CR 
energy spectrum is rather steep with $P\propto E^{-2.7}$ the majority of
the energy is deposited at lower energies with a peak at around a few
$GeV$, which is the expected range of significant dynamical impact of 
CRs on the ISM within a star forming galaxy.  The main acceleration
mechanism for galactic CRs, in particular those below the 'knee' in the CR
spectrum is considered to be \emph{diffusive shock acceleration} (DSA,
see e.g. \citealp{1978MNRAS.182..147B,1978ApJ...221L..29B}) and \emph{non-linear 
  DSA} \citep{2001RPPh...64..429M} in shocks of supernova remnants (SNR)
(see \citealp{2005JPhG...31R..95H}, for a review, and recent
observations by \citealp{2013Sci...339..807A}). Although both electrons 
and protons are accelerated in strong shocks, the protons carry most
of the energy stored as cosmic rays. 

The fraction of energy generated in supernova shocks is highly
uncertain and this process can, of course, not be simulated in galaxy
scale simulations. The estimates mostly range between 5 and 30 \% of 
the total supernova explosion energy
\citep{2006APh....25..246K,2010ApJ...712..287E} with increasingly
fundmagental, ab initio calculations now being made
(e.g. \citealp{2015ApJ...798L..28C}). If a significant  
fraction is 'stored' in cosmic rays - which cool much slower than
thermal gas by hadronic interactions and Coulomb and ionization losses
- they can be carried over large distances and significantly impact 
the ISM, provided the coupling between CRs and the thermal gas is
strong enough
(e.g. \citealp{1991A&A...245...79B,1997Natur.385..131Z}). CRs diffuse   
from the shocks and, later on, stream through the ISM. The bulk of CRs  
may be trapped at first by scattering at self-excited Alv\'en waves,
which causes a slowed down outward diffusion and additional mometum
transfer to the gas \citep{2015ApJ...798L..28C}. Once the cosmic rays
are able to escape the supernova remnant, a streaming instability will be excited if
the necessary conditions for the dynamical coupling between the CRs
and the gas are met (i.e. a large-scale CR gradient towards the galactic 
halo is required, \citealp{1969ApJ...156..445K}), effectively
transferring CR energy and momentum to the thermal gas. In addition
the CRs exert a pressure on the thermal gas by scattering off Alfv\'en
waves.  It is therefore expected that CRs have a significant impact on
the thermal and dynamical properties of the ISM. The propagation of
cosmic rays through the ISM is complex and often approximated by
diffusion with coefficients of the order $\kappa_{\mathrm{CR}} \propto
10^{28} - 10^{29} cm s^{-1}$
(e.g. \citealp{1998ApJ...509..212S,2011ApJ...729..106T,2013A&A...552A..19T}). Locally
the diffusion might be anisotropic with significantly smaller 
coefficients perpendicular to the magnetic field lines. On global
galactic scales of the Milky Way, however, the diffusion can be
considered isotropic \citep{2007ARNPS..57..285S}. The lifetime of the
several GeV cosmic ray fluid is known from radioactive dating to be
roughly 10 million years.   

In addition to thermal and radiation pressure caused by stellar
feedback, CRs turn out to be an important agent on galactic scales
and, once accelerated in regions of local feedback from star
formation, they might be efficient in supporting or even driving galactic
outflows.  Already in the beginning of the 90's it has
been proposed that the combined effect of thermal pressure, MHD waves,
and an effective (non-thermal) CR pressure \citep{1982A&A...116..191M}
is able to drive a galactic wind \citep{1991A&A...245...79B,1993A&A...269...54B},
even if the star formation rate is moderate. Recent observations of
the starburst galaxy M82 \citep{2009Natur.462..770V} reveal
CR densities which are 500 times higher than in the Milky Way and
thus clearly link the CR density with regions of highly efficient
star formation. Other groups have argued that the galactic
wind in M82 is purely driven by strong stellar feedback
\citep{1985Natur.317...44C,1996SSRv...75..279V}. However, in normal
spirals like the Milky Way, stellar feedback is probably not strong
enough to drive a large-scale  galactic wind. Nevertheless, recent
ROSAT observations of the Milky Way show extended, soft X-ray emission,
which is best explained with a kpc-scale wind for which the cosmic ray
pressure may be essential \citep{2008ApJ...674..258E}.  

In galaxy scale hydrodynamics simulations with SPH and AMR codes,
cosmic rays have recently been included as a separate fluid, as their
mean free path is shorter than the typical length scales resolved
\citep{1975MNRAS.172..557S}. The fluid is treated as a relativistic gas with   
$\gamma_{\mathrm{cr}} = 4/3$ and is advected with the gas. The
resulting total pressure is $p_{\mathrm{tot}} = p_{\mathrm{th}} +
p_{\mathrm{cr}}$ with $p_{\mathrm{cr}} = (\gamma_{\mathrm{cr}} -1) 
E_{\mathrm{cr}}$. In addition the cosmic rays are allowed to diffuse
through the ISM. This is treated approximately either by streaming
\citep{2012MNRAS.423.2374U} or by isotropic diffusion with a typical
diffusion coefficient 
\citep{2007A&A...473...41E,2008A&A...481...33J,2013ApJ...777L..16B,2014MNRAS.437.3312S}. 
In simulations following magnetic fields the diffusion is treated
anisotropically with one or two orders of magnitude lower diffusion
coefficients along the magnetic field lines
\citep{2012ApJ...761..185Y,2013ApJ...777L..38H,2016ApJ...824L..30P}.  

All simulations including cosmic rays on galactic scales indicate that
they can  significantly support the driving of bipolar galactic winds
with velocities exceeding the local escape speed and with mass loading
greater than unity. The winds are driven by the additional pressure
gradient due to cosmic rays in low density regions
\citep{2012MNRAS.423.2374U,2013ApJ...777L..38H,2013ApJ...777L..16B,2014MNRAS.437.3312S,2016ApJ...824L..30P,2016MNRAS.456..582S}. This
process is only efficient if cosmic rays can diffuse out of the high 
density regions, where they are dynamically unimportant, to build-up a
galaxy wide vertical gradient (see
e.g. \citealp{2012MNRAS.423.2374U,2013ApJ...777L..16B}). The mass
loading of cosmic ray driven winds is higher for lower mass galaxies
\citep{2011MNRAS.410.1975W,2012MNRAS.423.2374U,2013ApJ...777L..16B}
but also for gas rich massive galaxies the effect is significant
\citep{2013ApJ...777L..38H}. Cosmic ray driven outflows can also
support stable configurations of open magnetic field lines originating
from regions of high star formation rates \citep{2013ApJ...777L..38H}.   
These simulations might also be able to explain the detection of strong magnetic
fields at large radii ($\sim 50 kpc$) around star forming galaxies at
intermediate and high redshift \citep{2013ApJ...772L..28B}.

Recently, simulations of the impact of cosmics rays on the
interstellar medium on smaller scales have confirmed the ideas brought
forward by larger scale galaxy formation simulations and analytic
estimates. It was shown by \citet{2016ApJ...816L..19G}, using
magneto-hydrodynamic simulations with anisotropic diffusion that
cosmic rays indeed support the launching of outflows from galactic
disks and the basic results seems to be insensitive of the
magneto-hydrodynamical method used, details of the star formation
algorithm and the presence of self-gravity
\citep{2016ApJ...827L..29S}. It remains to be seen whether the
supporting role of cosmic rays (i.e. the pressure gradient) for the
driving of outflows is significant or whether the computations still
suffer from inaccuracies in capturing the accurate effect of supernova
explosions on the formation of a hot, wind driving, gas phase. 

The simulations presented above can only be the starting point for
more detailed investigations of the potential importance of cosmic
rays for the driving of galactic outflows. In galaxy scale
simulations, the galactic ISM is in general unresolved, with a 
significantly simplified treatment of energy injection by SNe as well as cosmic
ray transport. Also the global and local evolution of the galactic
magnetic field has to be considered and significant progress has been made recently
\citep{2005JCAP...01..009D,2008SSRv..134..311D,2009MNRAS.398.1678D,2009MNRAS.397..733K,2010ApJ...716.1438K,2010A&A...523A..72D,2013MNRAS.432..176P,2014ApJ...783L..20P,2016ApJ...824L..30P,2016MNRAS.457.1722R}. If  
it turns out that CRs and magnetic fields are indeed as
important as suggested they have the potential to become a 'global
player', regulating  the evolution of the ISM and the efficiency of
galaxy formation across cosmic times. 

\subsection{Mechanical and radiative AGN feedback} 
\label{mech}

While the significant work done by many groups using 'thermal
AGN feedback' (see Section \ref{blackholefeedback}) has solid motivation and has 
produced many quite useful results, the somewhat arbitrary physical
implementation has troubling aspects that several groups have
attempted to remedy. First, there is  no physical means specified for
communicating the energy from the black holes to the surrounding gas
in these treatments (see discussion in \citealp{2010ApJ...722..642O}). In actual AGN systems, 
there are high velocity winds and radiation output for both of which
there is a momentum associated with the energy transfer and there is a
spatial direction for the momentum outflow
\citep{2005MNRAS.363L..91C,2008MNRAS.388.1011M,2012ARA&A..50..455F}. But 
in a purely thermal feedback approach these physical factors are
missing and, more importantly, the mass to which the thermal feedback
energy is distributed is not specified or defined by a heating
threshold (Section \ref{blackholefeedback}). Adding this
energy to a very small mass would produce very high temperatures and
radiative energy losses, adding it to a very large mass would 
produce small additional velocities compared to the virial velocities
in the galaxy, so adding to the 'just right' amount of mass seems to
be required. But if teh added feedback energy is deposited with the
appropriate accompaning momentum, as observed in BAL winds, the energy
redistribution is achieved by physical means. 

There exists a characteristic electromagnetic spectrum for AGN with
peaks in the infrared, ultraviolet and X-ray
(e.g. \citealp{2004MNRAS.347..144S}) that can be taken as the emitted
output with the ratio of electromagnetic output to mass accretion
given by the Soltan argument \citep{ 1982MNRAS.200..115S}. In addition
to relativistic jets there are also broad-line winds observed to be commonly
emitted by AGN \citep{2014ARA&A..52..529Y} and these too have been
calibrated to the accretion rate empirically
(\citet{2013MNRAS.436.3286A}, see also
\citealp{1999agnc.book.....K}). Thus one can  specify the output in
mass, energy, momentum and  radiation in various bands per mass
accreted on the basis of approximate matches to observed AGN, and then
the thermal, mechanical and radiative coupling of these components to
the surrounding medium should be handled automatically by the hydro codes
being utilized. When taking this approach (and including the Eddington
force rather than putting in a limit) the accretion rate self-adjusts,
so no arbitrary multiplicative 'boost' factors may be needed in
implementing the Bondi-Lyttleton accretion. First attempts to
approximately include
the effects of UV and X-ray emission from accreting black holes have been made
(e.g. \citealp{2011ApJ...738...16H,2015MNRAS.449.4105C,2013MNRAS.436.3031V,2015ApJ...800...19R,2016arXiv160606281B}). 
Both mechanical (BAL) and radiative effects are included in
\citet{2016arXiv161009389C} for cosmological simulartions and in
\citet{2016MNRAS.458..816H} for high-resolution simulations of
galactic centers.  
However, in the end, the results of this more
complex approach are in fact similar to those of the thermal
feedback approach with regard to regulating the overall mass growth of
the central black hole, but other aspects are very different. For
example, the fluctuation level of the kinetic feedback is far more
extreme with 'on' periods rarely exceeding several million years and a
small overall duty cycle being expected. And, as noted earlier,
correct prediction of the thermal X-ray emssion from massive galaxies
provides a strong test of the feedback mechanism which indicates that
including a kinetic wind component is essential
\citep{2015MNRAS.449.4105C,2016arXiv160703486W}.

Jets from AGN are frequently observed and feedback from this
mechanism is sometimes called 'radio mode'
\citep{2012ARA&A..50..455F,2014ARA&A..52..589H}. Some cosmological galaxy evolution 
treatments have included this process (e.g. \citealp{2014MNRAS.444.1453D}), the
reality of which is not in doubt. But narrow relativistic jets
effectively drill through the ISM within galaxies, leaving a dramatic
imprint on the surrounding cluster gas but probably not communicating
significant amounts of energy or momentum to the ISM of the parent
galaxy (\citealp{2006ApJ...645...83V}). However, potential coupling mechanisms
e.g. turbulent mixing and dissipation exist and are studied in detail
with higher resolution simulations
\citep{2004MNRAS.348.1105O,2008ApJ...686..927S,2012MNRAS.424..190G,2013ApJ...768...11B,2014ApJ...789...54L}. Jets 
being extremely important in regulating cooling flows within  
clusters are probably less important than other, more prosaic
processes in determining the internal evolution of galaxies.

\section{Conclusion \& Outlook}

Since the advent of self-consistent cosmological numerical simulations
about 35 years ago significant progress in understanding galaxy
formation has been made. Modern super-computers and numerical
algorithms have become powerful enough to allow the simulation of
individual galaxies at relatively high resolution as well as the
evolution of galaxy populations in representative cosmological
volumes. With different numerical
techniques (smoothed particle hydrodynamics, meshless particle
hydrodynamics, adaptive mesh refinement
and moving mesh hydrodynamics, see review by
\citealp{2015ARA&A..53...51S})) it is now possible to simulate galaxy
populations (at spatial resolutions of 0.5 - 3 kpc) with a realistic 
cosmological evolution of sizes, abundances, star formation rates, dark matter
fractions, gas fractions as well as stellar and black hole masses,
from well defined initial conditions, for a direct comparison to
observational galaxy surveys
(e.g. \citealp{2013MNRAS.434.2645D,2014MNRAS.442.2304H,2015MNRAS.446..521S,2015MNRAS.452..575S,2015MNRAS.450.1349K}). 

Zoom simulations of individual galaxies (at resolutions $ < 500 pc$) allow
a detailed investigation of the formation processes and the
consequences for the internal galaxy structure for direct
comparison to high resolution observations
(e.g. \citealp{2011ApJ...742...76G,2013MNRAS.428..129S,2014MNRAS.445..175G,2014MNRAS.445..581H,2013MNRAS.433.3297D,2014MNRAS.437.1750M,2014MNRAS.441.3679A,2014MNRAS.444.3357N,2016ApJ...827L..23W}). Simulations of this kind (see also
\citealp{2015MNRAS.446.2038R,2015MNRAS.451...34R,2016MNRAS.458.3528H,2016Natur.535..523F,2016MNRAS.458..270R})
with high enough spatial resolution to represent the multi-phase
interstellar medium structure and stellar feedback more accurately
accounting for major effects like stellar winds, radiation and
supernovae have the potential to shed more light into the detailed
physical processes governing galaxy formation. 

The rapid recent progress can be considered a success in our quest for
a better understanding of galaxy formation.  It was mainly triggered
by the realization that thermal energy input from supernovae is most
likely insufficient to trigger outflows and 'galactic winds' with
other mechanisms being explored as the relevant drivers. These outflows, however,
play a major - if not the dominant - role for regulating the formation
of galaxies at low and high masses. They are most likely driven by
energy injection from newly formed stellar populations (cosmic rays, radiation and
winds, in addition to supernovae) and accreting black holes. As the outflows
are launched on parsec and sub-parsec scales, well below the resolution and physical
complexity limit of any cosmological simulation this has triggered a
wealth of sub-resolution models for stellar and AGN 
feedback (see Section \ref{stellarfeedback} and
\ref{blackholefeedback}). The fact that many models - even if  
conceptually very different - driving a 'reasonable galactic wind' can
'successfully' reproduce galaxy abundances and disk galaxy
morphologies (see e.g. Figs. \ref{mass_function} and \ref{disks}) indicates that the
essential characteristics of the problem has been disclosed. However,
the empirical nature of the subresolution models limit
the predictive power of the simulations and the literature becomes enriched
by parameter studies of particular implementations despite
obvious shortcomings of the respective models which are compensated by
adjusted parameters. 'Delayed cooling', 'stochastic thermal', and
'non-thermal' feedback models may significantly overestimate the
energy and momentum input into the ISM and the 'delay' time-scales 
are uncertain. 'Decoupled wind' models might not capture
(i.e. underestimate) the energy coupling to the local ISM and result
in unrealistic wind structures. Empirical 'momentum driving' models
rely on uncertain coupling efficiencies for
infrared radiation. Similarly, almost all AGN feedback models - on
cosmological scales - are of empirical nature with accretion and energy
conversion efficiencies adjusted, in a plausible fashion, to match
observed scaling relations.      

It will be a major theoretical challenge in theoretical galaxy
formation to understand stellar and AGN feedback in detail and
identify physically correct sub-resolution models taking into account
all relevant physical processes.  First promising steps in this
direction have been made from high resolution galaxy scale simulations
as well as simulations on smaller scales. A full accounting for the
energy input from stellar populations
(e.g. \citealp{2012MNRAS.421.3488H,2013ApJ...770...25A}),  the long
range effect of low and high energy radiation from stars and AGN
(e.g. \citealp{2012ApJ...754..125C,2013MNRAS.436.3031V,2014MNRAS.437.2882K,2015MNRAS.451...34R,2015ApJ...800...19R,2016arXiv160606281B}) 
and the consideration of other significant non-thermal components of
the ISM, namely magnetic fields and cosmic rays
(e.g.\citealp{2012MNRAS.423.2374U,2013ApJ...777L..38H,2013ApJ...777L..16B,2014MNRAS.437.3312S,2016ApJ...824L..30P})
are probably the most promising areas of galaxy formation research in
the future. These research directions mainly refer to number (2) of our
physics problem (Section \ref{intro}): the knowledge of the physical
processes primarily responsible for understanding each phase of
galactic evolution. Starting with well defined initial conditions we
can now roughly reproduce the scales and 
internal structures of common galaxy types and laboring with
increasing physical precision to correctly model the detailed
processes involved in feedback from stars and super-massive black
holes. 

\section*{Acknowledgements}

The authors acknowledge valuable input on the manuscript from 
O. Agertz, R. Dave, A. Dekel, Y. Dubois, P. Girichidis, M. Hanasz, M. Hirschmann,
P. Hopkins, N. Khandai, B. Moster, T. Peters, E. Puchwein, M. Rafieferantsoa, V. Springel, G. Stinson, R. Teyssier,
H. \"Ubler, M. Vogelsberger, and S. Zhukovska. The authors are
particularly grateful for the many detailed and valuable comments from
J. Kormendy, D. Nelson and J. Schaye.



\bibliographystyle{ARAstroBib}
\bibliography{./references.bib}

\end{document}